\documentclass[preprint]{aastex}
\usepackage{amsbsy}

\newcommand{\eqref}[1]{(\ref{#1})}
\renewcommand{\vec}[1]{\boldsymbol{#1}}

\shorttitle{Planetesimal Dynamics in a Turbulent Disk}
\shortauthors{Yang, Mac Low, \& Menou}

\begin{document}

\title{Planetesimal and Protoplanet Dynamics
       in a Turbulent Protoplanetary Disk:
       Ideal Stratified Disks}

\author{Chao-Chin Yang\altaffilmark{1}}
\affil{Department of Astronomy, University of Illinois, Urbana, IL~61801, USA}
\affil{Department of Astrophysics, American Museum of Natural History,
       New York, NY~10024, USA}
\email{ccyang@ucolick.org}

\author{Mordecai-Mark Mac Low}
\affil{Department of Astrophysics, American Museum of Natural History,
       New York, NY~10024, USA}
\email{mordecai@amnh.org}

\and

\author{Kristen Menou}
\affil{Department of Astronomy, Columbia University, New York, NY~10027, USA}
\affil{Perimeter Institute for Theoretical Physics, North Waterloo, Ontario, N2L~2Y5, Canada}
\email{kristen@astro.columbia.edu}

\altaffiltext{1}{Present address: Department of Astronomy and Astrophysics, University of California, Santa Cruz, CA~95064, USA.}

\begin{abstract}
Due to the gravitational influence of density fluctuations driven by magneto-rotational instability in the gas disk, planetesimals and protoplanets undergo diffusive radial migration as well as changes in other orbital properties. The magnitude of the effect on particle orbits can have important consequences for planet formation scenarios.  We use the local-shearing-box approximation to simulate an ideal, isothermal, magnetized gas disk with vertical density stratification and simultaneously evolve numerous massless particles moving under the gravitational field of the gas and the host star.  We measure the evolution of the particle orbital properties, including mean radius, eccentricity, inclination, and velocity dispersion, and its dependence on the disk properties and the particle initial conditions.  Although the results converge with resolution for fixed box dimensions, we find the response of the particles to the gravity of the turbulent gas correlates with the horizontal box size, up to 16 disk scale heights.  This correlation indicates that caution should be exercised when interpreting local-shearing-box models involving gravitational physics of magneto-rotational turbulence.  Based on heuristic arguments, nevertheless, the criterion $L_h / R \sim O(1)$, where $L_h$ is the horizontal box size and $R$ is the distance to the host star, is proposed to possibly circumvent this conundrum.  If this criterion holds, we can still conclude that magneto-rotational turbulence seems likely to be ineffective at driving either diffusive migration or collisional erosion under most circumstances.
\end{abstract}


%-------------------------------------------------------------------------------
\section{INTRODUCTION}

The gas disk surrounding a young stellar object is usually argued to be a turbulent accretion disk when the star is still in the process of accumulating materials \citep[see, e.g.,][]{FKR02}.  While the molecular viscosity of the gas is negligibly small, shear stress resulting from turbulence can effectively drive angular momentum transport in the disk.  At the same time, the dynamics of protoplanetary objects embedded in the disk is significantly affected by their interactions with the turbulent gas, at a stage when these objects are still amassing solids to grow in size \citep[e.g.,][and references therein]{PT06}.  Understanding particle dynamics in a turbulent gas disk, therefore, will provide insight into the paths along which new planets may eventually form.

One of the most promising mechanisms to drive the turbulence is through the magneto-rotational instability \cite[MRI; see, e.g.,][and references therein]{BH98}.  A weakly magnetized, differentially rotating gas disk is unstable to linear perturbations, and after nonlinear effects set in, the gas presumably reaches a statistically steady, sustained, turbulent state.  The underlying processes that determine the properties of this saturated state are still under active research \citep[e.g.,][]{JJK08,LLB09,eV09,mP10,OM11}.  Nevertheless, it is known that the strength of this magneto-rotational turbulence depends on the net magnetic flux threading the disk \citep[e.g.,][hereafter YMM09]{HGB95,JKH06,YMM09}, and thus can be controlled to the desired level and treated as a parameter.

The gravitational torques exerted on an embedded solid object due to density fluctuations in magneto-rotational turbulence have been shown to be stochastic, rendering the motion of the object a random walk (\citealt{LSA04,NP04,N05,OMM07}; YMM09; \citealt{NG10}) on top of the ordered migration expected in a laminar disk \citep{JGM06,AB09,KAB11}.  Note that the ordered migration may be either inward or outward in different regions of the disk \citep{PB10,PB11,LPM10}.  In addition to inducing this radial diffusive migration, the same gravitational force drives the diffusion of orbital eccentricity of the solid objects, increasing their velocity dispersion (\citealt{OIM07,IGM08}; YMM09; \citealt{NG10}).  The effect on the orbital inclination, however, has not been investigated yet, as that requires the inclusion of the vertical component of the gravity from the host star.  Overall, the gravitational influence of the turbulent gas on particle orbital properties may play an important role in the course of planet formation.

To assess the significance of this process in planet formation scenarios, numerical simulations simultaneously evolving magneto-rotational turbulence and particle dynamics can be conducted and the resulting particle orbital evolution can be statistically measured.  However, a consistent picture of the actual magnitude of this effect has been elusive.  Using the local-shearing-box simulations, we reported in YMM09 that the response of test particles to the gravity of the turbulent gas is systematically lower than what was previously reported in global disk models.  We also found in the same study that the effect significantly depends on the horizontal box size of the numerical models.  Recently, \citet{NG10} have found similar behavior in their local models and argued that large box size might be necessary to see convergence in the stochastic torque and forcing generated by the turbulence.

Extending the local unstratified disk models of YMM09, we now consider disks with vertical density stratification by including the linearized vertical gravity from the central star.  We examine whether a stratified disk model gives results consistent with a comparable unstratified model and whether convergence in particle orbital evolution, if any, can be achieved with local-shearing-box simulations.  In Section~\ref{S:model}, we describe in detail our numerical models.  We report the properties of our modeled magneto-rotational turbulence in Section~\ref{S:turbulence} and the statistical evolution of particle orbital properties in Section~\ref{S:particles}.  We analyze the gravitational field generated by the turbulence in Section~\ref{S:force} and discuss the convergence issues of our local models in Section~\ref{S:box_size}.  We conclude our discussion in Section~\ref{S:summary}.

%-------------------------------------------------------------------------------
\section{NUMERICAL MODELING} \label{S:model}

Directly extending the work we reported in YMM09, we continue to use the Pencil Code\footnote{The Pencil Code is publicly available at \texttt{http://code.google.com/p/pencil-code/}.} to model particles moving in magneto-rotational turbulence.  We describe in detail the equations and the relevant numerical techniques we implemented in the code.

%-------------------------------------------------------------------------------
  \subsection{Magnetohydrodynamics} \label{SS:mhd}
  
We use the local shearing box approximation \citep[e.g.,][]{GL65,BN95,HGB95}.  A local shearing box is a small Cartesian box at a large distance $R$ from the host star such that the center of the box revolves at the Keplerian angular speed $\Omega_K$.  The box is always oriented with the $x$-axis directed radially and the $y$-axis azimuthally.  In contrast to YMM09, we include the linearized vertical gravity from the host star so that the disk is vertically stratified.  We again impose a vertical, external magnetic field $\vec{B}_\mathrm{ext} = B_\mathrm{ext}\hat{\vec{z}}$ to maintain a non-zero magnetic flux.  The MHD equations then become
\begin{eqnarray} \label{E:mhd}
  \partial_t\rho
  - \frac{3}{2}\Omega_K x\partial_y\rho
  + \nabla\cdot(\rho\vec{u})
  &=& f_D,\label{E:mhd_cont}\\
  \partial_t\vec{u}
  - \frac{3}{2}\Omega_K x\partial_y\vec{u}
  + \vec{u}\cdot\nabla\vec{u}
  &=& -\frac{1}{\rho}\nabla p
  + \left(2\Omega_K u_y\hat{\vec{x}}
          - \frac{1}{2}\Omega_K u_x\hat{\vec{y}}
          - \Omega_K^2 z\hat{\vec{z}}\right)\nonumber\\
  &&+ \frac{1}{\rho}\vec{J}\times\left(\vec{B} + 
                                       \vec{B}_\mathrm{ext}\right)
  + \vec{f}_V,\label{E:mhd_mom}\\
  \partial_t\vec{A}
  - \frac{3}{2}\Omega_K x\partial_y\vec{A}
  &=& \frac{3}{2}\Omega_K A_y\hat{\vec{x}}
  + \vec{u}\times\left(\vec{B} + \vec{B}_\mathrm{ext}\right)
  + \vec{f}_R,\label{E:mhd_ind}
\end{eqnarray}
where $\rho$ is the gas density, $\vec{u}$ is the gas velocity relative to the background shear flow, $p$ is the gas pressure, $\vec{J} = \nabla\times\vec{B} / \mu_0$ is the electric current density, $\vec{B} = \nabla\times\vec{A}$, and $\mu_0$ is the vacuum permeability.  The terms $f_D$, $\vec{f}_V$, and $\vec{f}_R$ are numerical dissipation terms, including both hyper-diffusion and shock diffusion, that are needed to stabilize the scheme.  The reader is referred to YMM09 for their description.

To close the system of equations~\eqref{E:mhd_cont}--\eqref{E:mhd_ind}, we assume an isothermal ``equation of state'', $p = c_s^2\rho$, where $c_s$ is the isothermal speed of sound.  This is an ideal limit of the energy equation, in which heat diffusion is effectively instantaneous.  Gas buoyancy should thus be absent and the vertical mixing in the resulting magneto-rotational turbulence might be artificially enhanced \citep{BG09}.  Nevertheless, \citet{OM09} showed that vertical motions are not completely suppressed even in the case of adiabatic gas, where no heat diffusion occurs, and the differences in the mean kinetic and the magnetic energies of a vertically stratified gas disk between the isothermal and the adiabatic limits may only amount to about a factor of two.

We set up the gas density so that the gas is in vertical hydrostatic equilibrium initially:
\begin{equation} \label{E:rho_0}
  \rho_0(z) = \rho_m\exp\left(-\frac{z^2}{H^2}\right),
\end{equation}
where $\rho_m$ is the mid-plane gas density and $H = \sqrt{2}c_s / \Omega_K$ is the vertical disk scale height.\footnote{Note that there exists a factor of $\sqrt{2}$ difference between our definition of disk scale height $H$ (Equation~\eqref{E:rho_0}) and some other authors' in the literature.\label{scale_height}}  We fix the vertical dimension of the computational domain at $|z| \leq 2H$.  The initial magnetic vector potential $\vec{A}$ is set to zero, while the external magnetic field is fixed at such a level that the corresponding plasma $\beta_\mathrm{ext}(z) \equiv 2\mu_0 c_s^2\rho_0 / B_\mathrm{ext}^2$ is $\beta_\mathrm{ext}(0) = 6.2\times10^3$ in the mid-plane.  Gaussian noise of magnitude $10^{-3}\,H/P$ in each component of the gas velocity $\vec{u}$ is generated to seed the MRI, where $P = 2\pi / \Omega_K$ is the orbital period at the center of the shearing box.

Our assumption of a net vertical magnetic field allows us to compare our models directly to the unstratified models presented in YMM09.  The strength of this field controls the strength of the MRI-driven turbulence, so imposing it is an effective way to adjust the viscous stress generated by the MRI.  A similar procedure has been used in other recent publications including \citet{JO07,JYK09}, \citet{NG10}, \citet{SI09}, and \citet{SMI10}.  Some justification for this assumption can be found in the argument that at high altitude where the gas is tenuous, the magneto-convective motions tend to turn horizontal magnetic fields into vertical ones \citep[e.g.,][]{PW82,HT88,BN95}.  Therefore, we take the limit that $B_x = B_y = 0$ at the $z$-boundary while allowing the vertical component ($B_z$) to change freely.  In terms of the vector potential $\vec{A}$, this boundary condition translates to $\partial_z A_x = \partial_z A_y = A_z = 0$.  In the horizontal directions, we use the standard sheared periodic boundary conditions for the magnetic field, i.e., the field is periodic after a shift compensating for the Keplerian shear flow (\citealt{HGB95}; see also Section~\ref{SS:poisson}).

We explored several different vertical boundary conditions for the gas and the field.  We found that outflow boundary conditions for both led to numerical instability during the initial growth of the MRI.  Although \citet{SI09} and \citet{SMI10} did publish models using outflow boundary conditions, their disk lost significant amounts of mass through the boundary during their run.  We also found that periodic boundary conditions on both gas and field led to random crashes after tens of orbits of saturated turbulence for reasons that remain unclear.  On the other hand, the combination of the vertical field boundary condition with periodic boundary conditions in the gas leads to a stable numerical result for the hundreds of orbits required for us to accurately measure the evolution of particle orbital properties \citep[see also][]{PW82}.  The horizontal boundary conditions are again sheared periodic.  In this setup, the total mass is conserved while saturated turbulent motion can be maintained near the vertical boundaries at a steady level.

While it is interesting in itself to study the coronal structure and possibly the net flow of the gas \citep{SI09,SMI10,FD11}, we specifically exclude the modeling of the coronal regions to maintain the steadiness of the magneto-rotational turbulence and thus improve the statistical measurement of the particle orbital properties.  The gravitational acceleration due to the host star across the boundary does not present any numerical difficulty.  Even though the vertical component of the gravitational acceleration changes sign when a fluid element is wrapped around the vertical boundary, the vertical velocity component of the element retains sign such that its velocity and acceleration always have the same sign after reemerging from the opposite boundary, resembling material ballistically falling back down onto the disk.  Numerically, no z-derivative of the gravitational acceleration due to the host star is ever evaluated (see Equation~\eqref{E:mhd_mom}), so the discontinuity of the acceleration across the vertical boundary is not an issue here.  Turbulence properties near the disk mid-plane, where the gravitational influence of the turbulent gas on the particles is strongest \citep{OMM07}, appear insensitive to the choice of vertical boundary conditions \citep{JYK09,DSP10,GG11}, as we also demonstrate in Section~\ref{SS:bce}.

For large box simulations, concerns have been raised that artificial numerical diffusion varying with $x$ may occur due to the radial dependence of the shear velocity and thus corresponding truncation errors on a fixed grid \citep{JGG08}.  Explicit treatment of the shear advection may be needed to eliminate these truncation errors.  In the Pencil Code, such a technique, dubbed shear advection by Fourier interpolation (SAFI), has been implemented \citep{JYK09}.  In our work, we implement SAFI for all of our simulations to remove this undesirable numerical effect.

%-------------------------------------------------------------------------------
  \subsection{Poisson Equation with Isolated Boundary Condition in $z$-direction} \label{SS:poisson}

To find the gravitational influence of the turbulent gas on the movement of solid particles, we need to solve the Poisson equation for the fluctuation potential $\Phi_1$:
\begin{equation} \label{E:poisson}
  \nabla^2\Phi_1 = 4\pi G\rho_1,
\end{equation}
where $G$ is the gravitational constant and $\rho_1(x,y,z) \equiv \rho(x,y,z) - \rho_0(z)$ is the density fluctuation with respect to the basic state of the gas stratification $\rho_0(z)$, which is given by Equation~\eqref{E:rho_0}.  Note that we have neglected the self-gravity of the gas in Equation~\eqref{E:mhd_mom} on the assumption that the disk is gravitationally stable (see YMM09), so the solution to Equation~\eqref{E:poisson} does not affect the gas dynamics.

Although the vertical boundary conditions for the gas density are assumed to be periodic in the MHD equations (see Section~\ref{SS:mhd}), we find it inappropriate to assume the same in solving the Poisson equation~\eqref{E:poisson}.  Consider, for example, density perturbations near the top edge of the computational domain.  If periodic boundary conditions in the vertical direction were adopted, particles near the mid-plane would experience the gravity of the same density perturbations at almost equal distance from above and below, resulting in cancellation of the gravitational force exerted by these perturbations on the particles.  This would amount to undesirable, potentially large, numerical errors for calculating the gravity of density perturbations at high altitude.

Therefore, we instead adopt isolated or open boundary conditions in the vertical direction to solve Equation~\eqref{E:poisson}.  With these boundary conditions, we assume that there are no fluctuations in density outside the vertical boundary, i.e., $\rho_1(x,y,z) = 0$ for $|z| > 2H$.  The boundary conditions in the horizontal directions remain sheared periodic.

To implement approximate isolated vertical boundary conditions with minimal computational cost, we use a variation of the \citet{HE88} fast algorithm to solve the Poisson equation~\eqref{E:poisson}, combining it with the Fourier interpolation technique to achieve sheared periodicity \citep{BI07,JO07}.  The essence of this algorithm is to double the computational domain in the $z$-direction by appending regions with zero density fluctuation\footnote{In our case, we append $\rho_1(x,y,z) = 0$ for $2H < z < 6H$ so the vertical domain now covers $-2H < z < 6H$.} and apply fast Fourier transforms to this expanded domain.  Although the vertical boundary conditions remain periodic with Fourier transforms, the $1/r^2$ nature of gravity means that influence from the periodic copies, now at least $2 L_z$ away, of the original computational domain of size $L_z$ onto itself is significantly reduced.

We now describe the algorithm to solve Equation~\eqref{E:poisson} in detail.  After expanding the vertical dimension of the computational domain and initializing the expanded region with zero density fluctuations, $\rho_1(x,y,z)$ is Fourier transformed in the $y$-direction into $\hat{\rho}_1(x,k_y,z)$, where $k_y$ is the $y$-wavenumber.  The result is phase shifted to recover periodicity in $x$-direction: $\check{\rho}_1(x,k_y,z) = \hat{\rho}_1(x,k_y,z)\exp\left[-i(3/2)\Omega_K k_y x\delta t \right]$, where $\delta t$ is the time step.  Then $\check{\rho}_1(x,k_y,z)$ is Fourier transformed in the $x$-direction into $\tilde{\rho}_1(k_x,k_y,z)$, where $k_x$ is the $x$-wavenumber.  For the modes $k_x^2 + k_y^2 > 0$, $\tilde{\rho}_1(k_x,k_y,z)$ is Fourier transformed in $z$-direction to find $\breve{\rho}_1(k_x,k_y,k_z)$, where $k_z$ is the $z$-wavenumber.  The fluctuation potential in Fourier space $\breve{\Phi}_1(k_x,k_y,k_z)$ is calculated using the usual convolution theorem:
\begin{equation} \label{E:convol}
  \breve{\Phi}_1(k_x,k_y,k_z)
  = -\frac{4\pi G}{\left(k_x + 3\Omega_K k_y\delta t / 2\right)^2 +
                   k_y^2 + k_z^2}
     \breve{\rho}_1(k_x,k_y,k_z),\quad
  \textrm{for}\ k_x^2 + k_y^2 > 0.
\end{equation}
It is then inverse Fourier transformed in $k_z$ to find $\tilde{\Phi}_1(k_x,k_y,z)$.

In principle, the mode $k_x = k_y = 0$ can similarly be treated by Fourier transforms in the vertical direction and convolved by Equation~\eqref{E:convol} as above.  However, this mode has an analytical solution that we can employ with marginal computational cost to improve the solution to Equation~\eqref{E:poisson}.\footnote{We note that use of Equation~\eqref{E:convol} would yield a 50\% relative error in the vertical acceleration due to a horizontal uniform slab at a distance $L_z / 2$ away (after domain doubling) when compared to the exact acceleration derived from Equation~\eqref{E:sheet}.}  It represents a horizontally infinite, vertically thin mass layer of constant density at a given altitude $z$, where the density is equal to the horizontal average of the density fluctuation at $z$.  (The layer can be considered as either a slab of finite size $\Delta z$ and constant volume density $\tilde{\rho}_1(k_x=0,k_y=0,z)$ or as an infinitely thin sheet of surface density $\tilde{\rho}_1(k_x=0,k_y=0,z) \Delta z$; the solutions outside the layer are the same by symmetry and Gauss' law.)  The gravitational acceleration due to this mode is constant above and below $z$.  Therefore, the resultant fluctuation potential can be calculated by
\begin{equation} \label{E:sheet}
  \tilde{\Phi}_1(k_x=0,k_y=0,z) = 2\pi G\int\,\tilde{\rho}_1(k_x=0,k_y=0,z')|z - z'|dz',
\end{equation}
where we have arbitrarily defined the reference potential to be zero at each altitude.  The discretized version of Equation~\eqref{E:sheet} reads
\begin{equation} \label{E:dis_sheet}
  \tilde{\Phi}_1(k_x=0,k_y=0,z_j) =
  2\pi G\sum_k \tilde{\rho}_1(k_x=0,k_y=0,z_k)|j - k|\Delta z^2.
\end{equation}
Equation~\eqref{E:dis_sheet} remains exact for this mode given that we do not consider reconstruction of the density field within each cell.  Finally, we reverse the process to inverse Fourier transform $\tilde{\Phi}_1(k_x,k_y,z)$ in $k_x$ and $k_y$ to derive the fluctuation potential in real space $\Phi_1(x,y,z)$, incorporating the corresponding phase shifts with opposite sign.

In practice, it is not necessary to allocate storage space for the whole extended domain in the $z$-direction.  The appended zero density is only involved in the calculation between the forward and inverse Fourier transforms in $x$ and $y$.  In addition, the convolutions in $z$ (Equations~\eqref{E:convol} and~\eqref{E:dis_sheet}) for different modes $(k_x,k_y)$ are independent of each other.  Therefore, these convolutions can be distributed among processors and performed sequentially by allocating only a single one-dimensional working array along the $z$-direction.

%-------------------------------------------------------------------------------
  \subsection{Particle Dynamics} \label{SS:pd}

We continue to use the zero-mass approximation to model our solid particles as in YMM09.  In this approximation, particles behaves as test particles and only respond to the gravity of the host star and the gas.  We ignore the viscous drag forces between particles and gas.  This remains a good approximation for kilometer-sized planetesimals \citep[e.g.,][]{OMM07}.  We also ignore mutual gravitational interactions between particles, which helps us isolate the net effect induced by hydromagnetic turbulence.

The equations of motion for each particle, therefore, become
\begin{mathletters} \label{E:eomp}
  \begin{eqnarray}
    \frac{d\mathbf{x}_p}{dt}
      &=& \mathbf{u}_p
          - \frac{3}{2}\Omega_K x_p\hat{\mathbf{y}},\label{E:eomp1}\\
    \frac{d\mathbf{u}_p}{dt}
      &=& \left(2\Omega_K u_{p,y}\hat{\mathbf{x}}
                - \frac{1}{2}\Omega_K u_{p,x}\hat{\mathbf{y}}
                - \Omega_K^2 z_p\hat{\mathbf{z}}\right)
          + \mathbf{g}_0 - \nabla\Phi_1.\label{E:eomp2}
  \end{eqnarray}
\end{mathletters}
The vector $\mathbf{x}_p = \left(x_p, y_p, z_p\right)$ is the position of the
particle in the shearing box, while $\mathbf{u}_p = \left(u_{p,x}, u_{p,y}, u_{p,z}\right)$ is the velocity of the particle relative to the background shear flow.  In Equation~\eqref{E:eomp2}, the terms inside the parentheses stem from the linearized gravity of the host star and the Coriolis and centrifugal forces in the co-rotating frame of the shearing box.  The remaining terms on the right-hand side represent the acceleration attributed to the gas: $\mathbf{g}_0$ is the gravitational acceleration due to the basic state of the gas stratification and has the analytical expression
\begin{equation} \label{E:g_0}
  \mathbf{g}_0(z) = -2\pi^{3/2}G\rho_m H
                    \,\mathrm{erf}\left(\frac{z}{H}\right)
                    \hat{\mathbf{z}},
\end{equation}
while $\Phi_1$ is the gravitational potential due to density fluctuations with respect to the basic state and is the solution of the Poisson Equation~\eqref{E:poisson} (Section~\ref{SS:poisson}).  Note that by separating the density perturbation from the equilibrium density stratification and treating the gravity of the equilibrium state exactly, we improve the accuracy of the gravitational potential resulting from the turbulence.

The position $\mathbf{x}_p$ and velocity $\mathbf{u}_p$ of each particle is evolved in time by integrating the equations of motion~\eqref{E:eomp} simultaneously with the third-order Runge-Kutta steps advancing the MHD equations.  In addition to the Courant conditions set by the MHD equations, the time-step is limited by the absolute maximum of velocity $\vec{u}_p$ so that no particles can cross more than half a grid zone each timestep.  We calculate the gradient of the fluctuation potential $\nabla\Phi_1$ on the grid and then quadratically interpolate it onto each particle.

We uniformly distribute $128^2$ particles in a horizontal plane.  We do not allow these particles to move until after 20 orbital periods, when the turbulence has reached a statistically steady state.  The particles can have an initial eccentricity $e_0$ by setting an initial velocity $\vec{u}_p$ so that they are at the apogee of their orbits (see YMM09).  They can also have an initial inclination $i_0$ by placing the initial particle plane at a distance to the mid-plane.  Particles are wrapped around when crossing any of the six boundary planes of the shearing box.

%-------------------------------------------------------------------------------
\section{PROPERTIES OF MAGNETO-ROTATIONAL TURBULENCE} \label{S:turbulence}

In this section, we present several properties of the magneto-rotational turbulence in our numerical models and discuss their convergence with the box size and the resolution.

\begin{figure}[htbp]
\begin{center}
\plotone{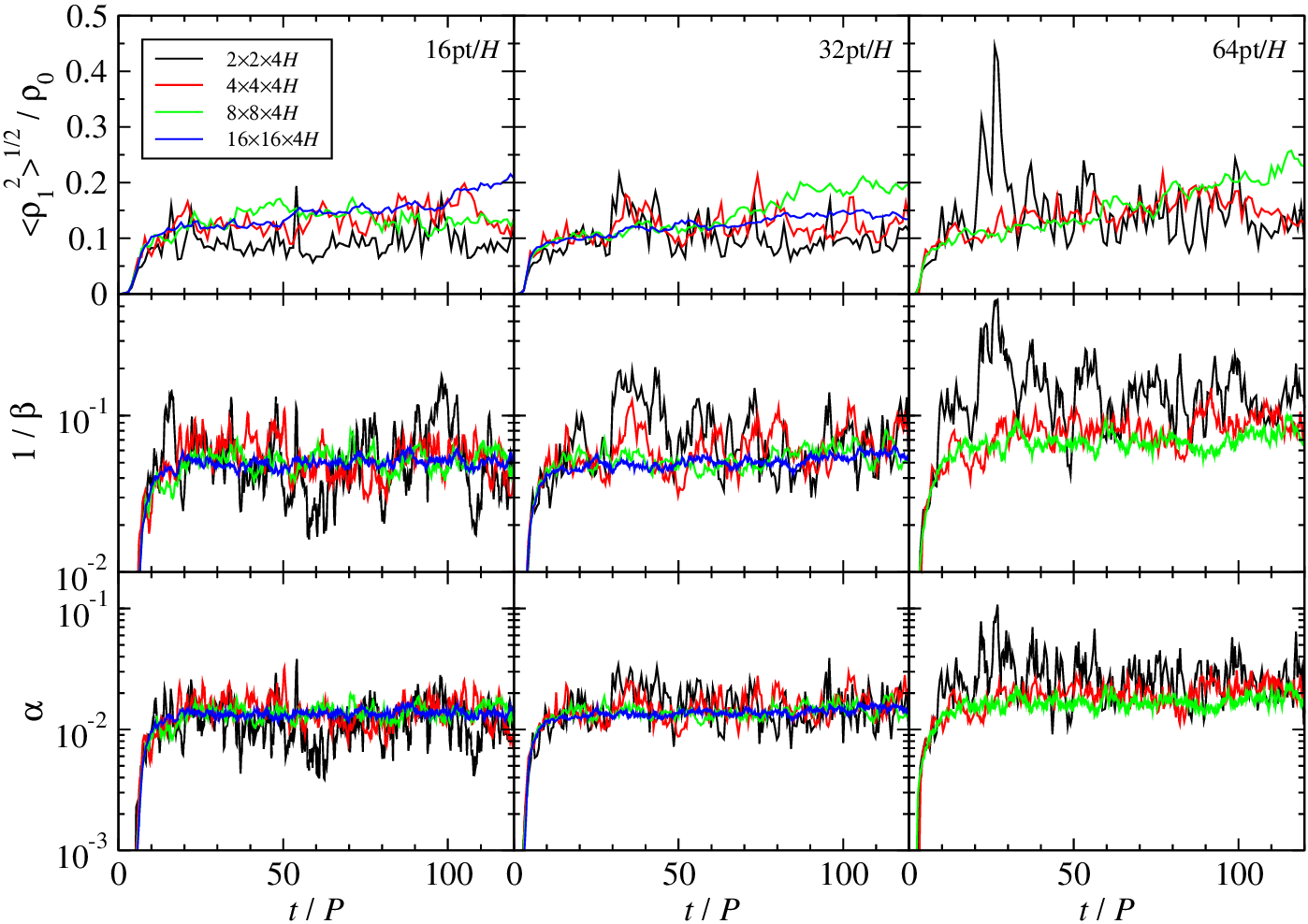}
\caption{Horizontally averaged properties of the modeled magneto-rotational turbulence in the mid-plane as a function of time.  The rows of panels from top to bottom are rms density fluctuation, inverse plasma $\beta$, and $\alpha$-parameter, respectively.  The columns are arranged with increasing resolution from left to right.  Lines of different colors denote measurements from boxes with different dimensions.}
\label{F:mri_t}
\end{center}
\end{figure}

Figure~\ref{F:mri_t} shows the horizontally averaged density fluctuation, inverse plasma $\beta$, and $\alpha$ parameter in the mid-plane ($z = 0$) as a function of time, for resolutions up to 64~points per $H$ and horizontal box dimensions up to $L_x = L_y = 16~H$.  The magnitude of the density fluctuation is indicated by the rms value relative to the equilibrium density $\langle\rho_1^2\rangle^{1/2} / \rho_0$, the plasma $\beta \equiv 2\mu_0\rho_0 c_s^2 / \langle\left|\vec{B} + \vec{B}_\mathrm{ext}\right|^2\rangle$ is the ratio of the equilibrium thermal pressure to the magnetic pressure, and $\alpha$ is calculated by \citep[e.g.,][]{aB98}
\begin{equation}
  \alpha = \frac{\sqrt{2}}{3}
           \frac{\langle\rho u_x u_y - B_x B_y / \mu_0\rangle}{\rho_0 c_s^2},
\end{equation}
which includes both Reynolds and magnetic stresses.\footnote{The bracket $\langle\ \rangle$ denotes the horizontal average of the enclosed variable at any given altitude $z$ and thus is a function of $z$.}
\placefigure{F:mri_t}

The magneto-rotational turbulence in the mid-plane of our models reaches a statistically steady state at about $t = 20P$.  In this saturated state, we see in general that the larger the horizontal size of the box, the smaller the amplitude of oscillation in the turbulence properties.  Little difference exists between an 8$\times$8$\times$4$H$ box and a 16$\times$16$\times$4$H$ box at the same resolution, indicating convergence with horizontal box dimension.  On the other hand, the amplitude of oscillation at saturation level increases with resolution for fixed box dimensions.  There exists some evidence that the magnetic activity and thus the $\alpha$ value increase with resolution, especially for the case of 2$\times$2$\times$4$H$ box, but this effect seems to become smaller with larger box sizes.  We also note that there exists a curious trend of increase in density perturbation with time for the case of an 8$\times$8$\times$4$H$ box at the resolution of 64~points/$H$.  In spite of these trends, the time average of each turbulence properties remains roughly the same for different resolutions and thus shows convergence.  We note that the turbulence properties at the saturated state in the mid-plane of our vertically stratified disks agrees with those in our unstratified disks of YMM09 \citep[see also][]{GG11}.  The means and standard deviations of these properties for each set of resolution and box dimensions are reported in Table~\ref{T:mri}, which includes the magnetic stress and the Reynolds stress as well.
\placetable{T:mri}

\begin{figure}[htbp]
\begin{center}
\epsscale{.8}
\plotone{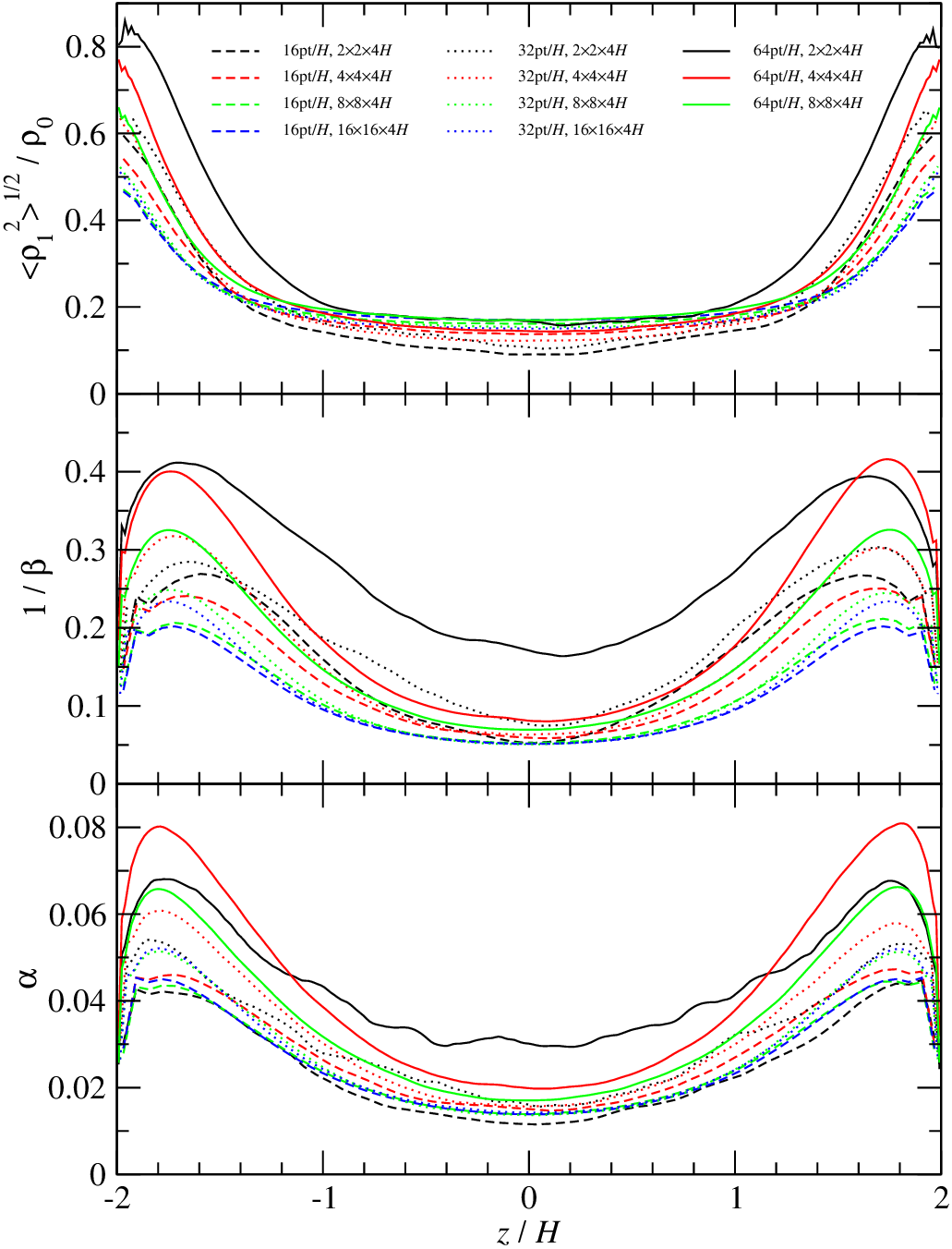}
\caption{Time-averaged vertical profiles of the modeled magneto-rotational turbulence at saturation stage for different resolutions and box dimensions.  The panels from top to bottom are rms density fluctuations, inverse plasma $\beta$, and $\alpha$ parameter, respectively.}
\label{F:mri_z}
\epsscale{1}
\end{center}
\end{figure}

Figure~\ref{F:mri_z} shows the time-averaged vertical profiles of density perturbation, inverse plasma $\beta$, and $\alpha$ parameter at saturation level over a period of $100P$.  Each of the three properties increases with $|z|$, indicating stronger turbulence at higher altitude.  The increasing activity with altitude is related to the increasing importance of the external magnetic field, which can be quantified by $\beta_\mathrm{ext}(z)$ (Section~\ref{SS:mhd}).  This agrees with the trend found in models of unstratified disks (\citealt{HGB95,JKH06}; YMM09).  Notice that $\beta_\mathrm{ext}(\pm2H) = 1.1\times10^2 \gg 1$ at our highest altitude, and thus the corona is not modeled in our simulations.  Given that our vertical height only amounts to 2$H$, the gas flow in our computational domain remains marginally stable against magnetic buoyancy during all stages of the development of the MRI \citep[e.g.,][]{SKH10,GG11,SHB11}.  As shown in Figure~\ref{F:mri_z}, the time-averaged inverse plasma beta near the vertical boundary amounts to only about 0.4, and thus on average, thermal pressure still dominates magnetic pressure there.
\placefigure{F:mri_z}

%-------------------------------------------------------------------------------
\section{ORBITAL PROPERTIES OF MASSLESS PARTICLES} \label{S:particles}

We now report the response of zero-mass particles to the gravity of the density fluctuations of the statistically steady, numerically convergent magneto-rotational turbulence described in Section~\ref{S:turbulence}.  The reference time $t = 0$ in the following discussion is the time at which the turbulent gas reaches its saturated state and the particles are allowed to move.\footnote{We choose this reference time to be $t = 20P$ as reported in Section~\ref{S:turbulence}.  See Figure~\ref{F:mri_t}.}
 
%-------------------------------------------------------------------------------
\subsection{Mean Orbital Radius} \label{SS:radius}

The evolution of the mean orbital radius of one particle can be found by averaging the radial position $x$ of the particle over each epicycle motion.  For the case of ideal unstratified disks, the distribution of particles in terms of the orbital radius change can be described by a time-dependent normal distribution centered at zero:
\begin{equation} \label{E:dist_dx}
f(\Delta x, t) = \frac{1}{\sqrt{2\pi}\sigma_x(t)}
                 \exp\left[-\frac{\Delta x^2}{2\sigma_x^2(t)}\right],
\end{equation}
where $\Delta x$ is the orbital radius change from the initial radius $x_0$ at $t = 0$ and $\sigma_x(t)$ is the time-dependent standard deviation.  We reported in YMM09 that $\sigma_x(t)$ depends on the properties of the gas disk and can be concisely expressed by
\begin{equation} \label{E:sigma_x}
\sigma_x(t) = C_x\xi H\left(\frac{t}{P}\right)^{1/2},
\end{equation}
where $C_x$ is a dimensionless proportionality constant\footnote{$C_x$ as well as other dimensionless constants introduced in the following discussions may depend on the $\alpha$ parameter.  This dependency is not investigated in this work.} and $\xi \equiv 4\pi G\rho_0 P^2$ is a dimensionless quantity indicating the strength of the gas gravity.  Equations~\eqref{E:dist_dx} and \eqref{E:sigma_x} demonstrate the diffusive nature of particle radial migration driven by magneto-rotational turbulence.

For the case of ideal stratified disks studied here, we again find the evolution of particle orbital radius can be described by Equations~\eqref{E:dist_dx} and \eqref{E:sigma_x}.  The dimensionless quantity $\xi$ is now defined by
\begin{equation}
\xi \equiv 4\pi G\rho_m P^2 = 4(2\pi)^{3/2} / Q_g
\end{equation}
in terms of the mid-plane gas density $\rho_m$, where $Q_g$ is the Toomre stability parameter for the gas.  In Table~\ref{T:orbit}, we report the measured values of the constant $C_x$ from our simulations at different resolutions and horizontal box sizes for the case of particles with zero initial inclination.  Comparing the value from the 2$\times$2$\times$4$H$ stratified model at a resolution of 64~points/$H$ with that from the unstratified model at the same resolution and horizontal box size (see Equation~(15) of YMM09), we find excellent agreement in the two $C_x$ values.

\begin{figure}[htbp]
\begin{center}
\plotone{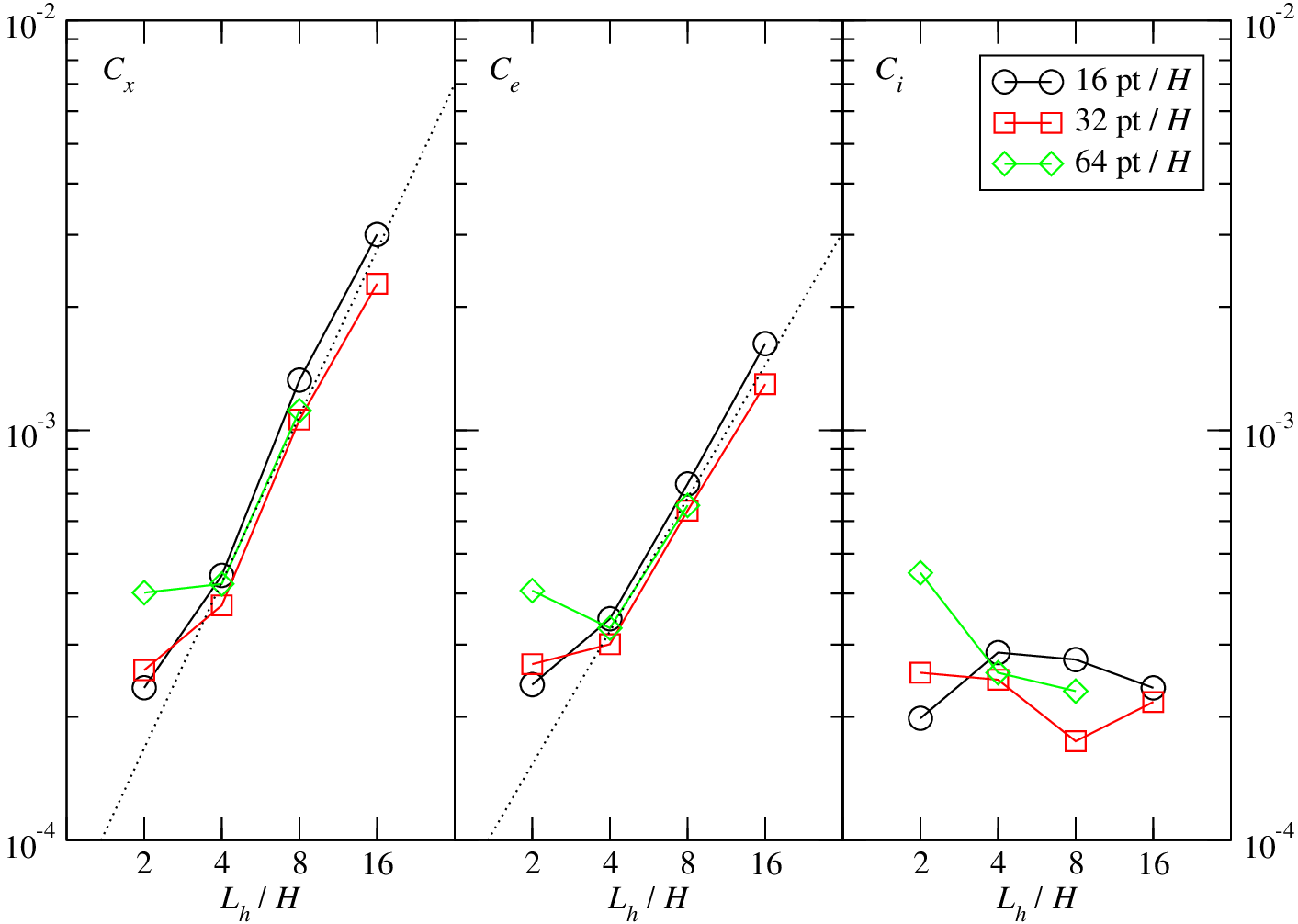}
\caption{Dimensionless proportionality constants $C_x$, $C_e$, and $C_i$ as a function of horizontal box size $L_h = L_x = L_y$.  They indicate the strength of diffusion in mean orbital radius, eccentricity, and inclination of massless particles moving in magneto-rotational turbulence and are defined in Equations~\eqref{E:sigma_x}, \eqref{E:sigma_e}, and~\eqref{E:sigma_i}, respectively.  The dotted lines are power-law fits to data points with $L_h \geq 4H$.}
\label{F:orbit}
\end{center}
\end{figure}

As can be seen from Table~\ref{T:orbit} and Figure~\ref{F:orbit}, the value of $C_x$ remains roughly the same with different resolutions for fixed box dimensions (except the anomaly shown by the 2$\times$2$\times$4$H$ box at a resolution of 64~points/$H$).  Conversely, it is significantly dependent on the horizontal box size $L_h \equiv L_x = L_y$.  The larger the size, the stronger the influence of the turbulence on the particle orbital radius.  This relationship can be represented by the following power-law fit:
\begin{equation} \label{E:cx}
C_x \simeq 6.6\times10^{-5}\,(L_h / H)^{1.35}.
\end{equation}
We find no evidence of convergence with horizontal box size up to $L_h = 16H$, the largest size we have investigated.
\placefigure{F:orbit}

\begin{figure}[htbp]
\begin{center}
\plotone{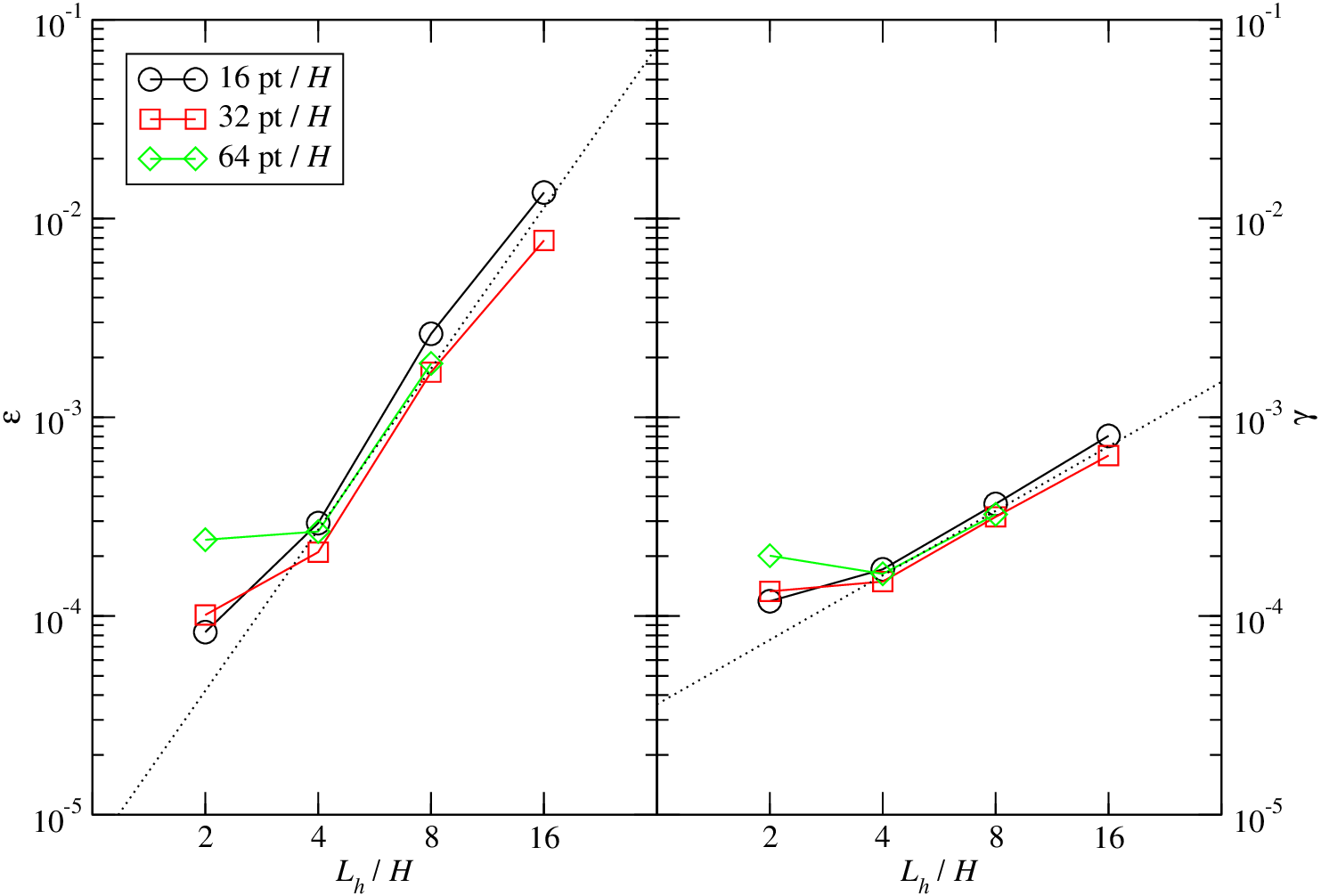}
\caption{Dimensionless constants $\epsilon$ and $\gamma$ as a function of horizontal box size $L_h$.  The constant $\epsilon$ indicates the strength of diffusive migration driven by turbulence and was defined by \citet{JGM06}, while $\gamma$ is related to the strength of eccentricity excitations (when $e_0 = 0$) due to turbulence and was defined by \citet{IGM08}.  The dotted lines are power-law fits to data points with $L_h \geq 4H$.}
\label{F:epsilon_gamma}
\end{center}
\end{figure}

Equation~\eqref{E:sigma_x} can be transformed into the diffusion coefficient $D(J)$ introduced by \citet{JGM06} for describing the radial random walks of orbiting particles induced by turbulent torques, in terms of the Keplerian orbital angular momentum $J$.  The reader is referred to YMM09 for a detailed description of this procedure.  We emphasize here that this transformation involves no assumption about the correlation time of the stochastic torques and is thus a direct measurement of $D(J)$.  Using a heuristic choice of dimension for the diffusion coefficient, \citet{JGM06} defined a dimensionless parameter $\epsilon$ to represent the magnitude of $D(J)$.  We report our measured values of $\epsilon$ in Table~\ref{T:orbit}.  The dependence of $\epsilon$ on the horizontal box size $L_h$ in our models for $L_h \gtrsim 4H$ is shown in Figure~\ref{F:epsilon_gamma} and can be written as
\begin{equation} \label{E:epsilon}
\epsilon \simeq 6.5\times10^{-6}\,(L_h / H)^{2.69}.
\end{equation}
\placefigure{F:epsilon_gamma}

%-------------------------------------------------------------------------------
\subsection{Eccentricity} \label{SS:ecc}

The amplitude of each epicyclic oscillation of a particle gives the instantaneous eccentricity of the particle orbit.  We reported in YMM09 that in ideal unstratified disks, the distribution of particles in terms of the eccentricity deviation should be a time-dependent normal distribution centered at zero, as long as the particles have non-negligible initial eccentricity (c.f.\ Equation~\eqref{E:dist_dx}):
\begin{equation} \label{E:dist_de}
f(\Delta e, t) = \frac{1}{\sqrt{2\pi}\sigma_e(t)}
                 \exp\left[-\frac{\Delta e^2}{2\sigma_e^2(t)}\right],
\end{equation}
where $\Delta e$ is the eccentricity deviation from the initial eccentricity $e_0$ at $t = 0$ and the time-dependent standard deviation $\sigma_e(t)$ can be written as
\begin{equation} \label{E:sigma_e}
\sigma_e(t) = C_e\xi\left(\frac{H}{R}\right)\left(\frac{t}{P}\right)^{1/2},
\end{equation}
where $C_e$ is a dimensionless proportionality constant.  We emphasize that the time-dependent Rayleigh distribution\footnote{Note that the standard deviation of a Rayleigh distribution is equal to $\sigma_e\sqrt{(4-\pi)/2}$.} found for particles with zero (or negligible) initial eccentricity
\begin{equation} \label{E:dist_e}
f(e,t) = \frac{e}{\sigma_e^2(t)}\exp\left[-\frac{e^2}{2\sigma_e^2(t)}\right]
\end{equation}
is a manifestation of Equation~\eqref{E:dist_de} since the eccentricity $e$ is a positive definite quantity.  Equations~\eqref{E:dist_de} and \eqref{E:dist_e} share the same time-dependent parameter $\sigma_e(t)$, and there exists no evidence that $\sigma_e(t)$ depends on the initial eccentricity $e_0$.

We find that the same evolution of particle distribution in eccentricity deviation also holds for the case of ideal stratified disks.  The measured values of the constant $C_e$ for different resolutions and horizontal box sizes when particles have zero initial inclination are listed in Table~\ref{T:orbit}.  The same comparison performed in Section~\ref{SS:radius} for the orbital radius evolution indicates that the eccentricity evolution in a stratified disk again agrees very well with that in an unstratified disk (see Equation~(17) of YMM09).  These comparisons demonstrate that a local unstratified disk model gives results consistent with the mid-plane of a local stratified disk model that has the same physical conditions except vertical stratification.

As in the case of orbital radius discussed in Section~\ref{SS:radius}, the eccentricity evolution of the particles does not noticeably depend on the resolution for given box dimensions, while it is strongly affected by the horizontal box size (Table~\ref{T:orbit} and Figure~\ref{F:orbit}).  A power-law regression gives
\begin{equation} \label{E:ce}
C_e \simeq 7.2\times10^{-5}\,(L_h / H)^{1.08},
\end{equation}
which is close to a linear relation to the horizontal box size.  We discuss the box-size effect on both the orbital radius and the eccentricity in Section~\ref{S:box_size}.

The eccentricity driven by magneto-rotational turbulence enhances orbital crossing among planetesimals and thus increases the chance of collisions between them.  \citet{IGM08} defined a dimensionless parameter $\gamma$ to represent the strength of this effect.  Equation~\eqref{E:sigma_e} can be used to estimate the value of $\gamma$, and the reader is referred to YMM09 for a detailed description of this procedure.  We report in Table~\ref{T:orbit} our measured values of $\gamma$.  As shown in Figure~\ref{F:epsilon_gamma}, the dependence of $\gamma$ on the horizontal box size $L_h$ in our models for $L_h \gtrsim 4H$ can be described by
\begin{equation} \label{E:gamma}
\gamma \simeq 3.6\times10^{-5}\,(L_h / H)^{1.08}.
\end{equation}

%-------------------------------------------------------------------------------
\subsection{Inclination}

\begin{figure}[htbp]
\begin{center}
\plotone{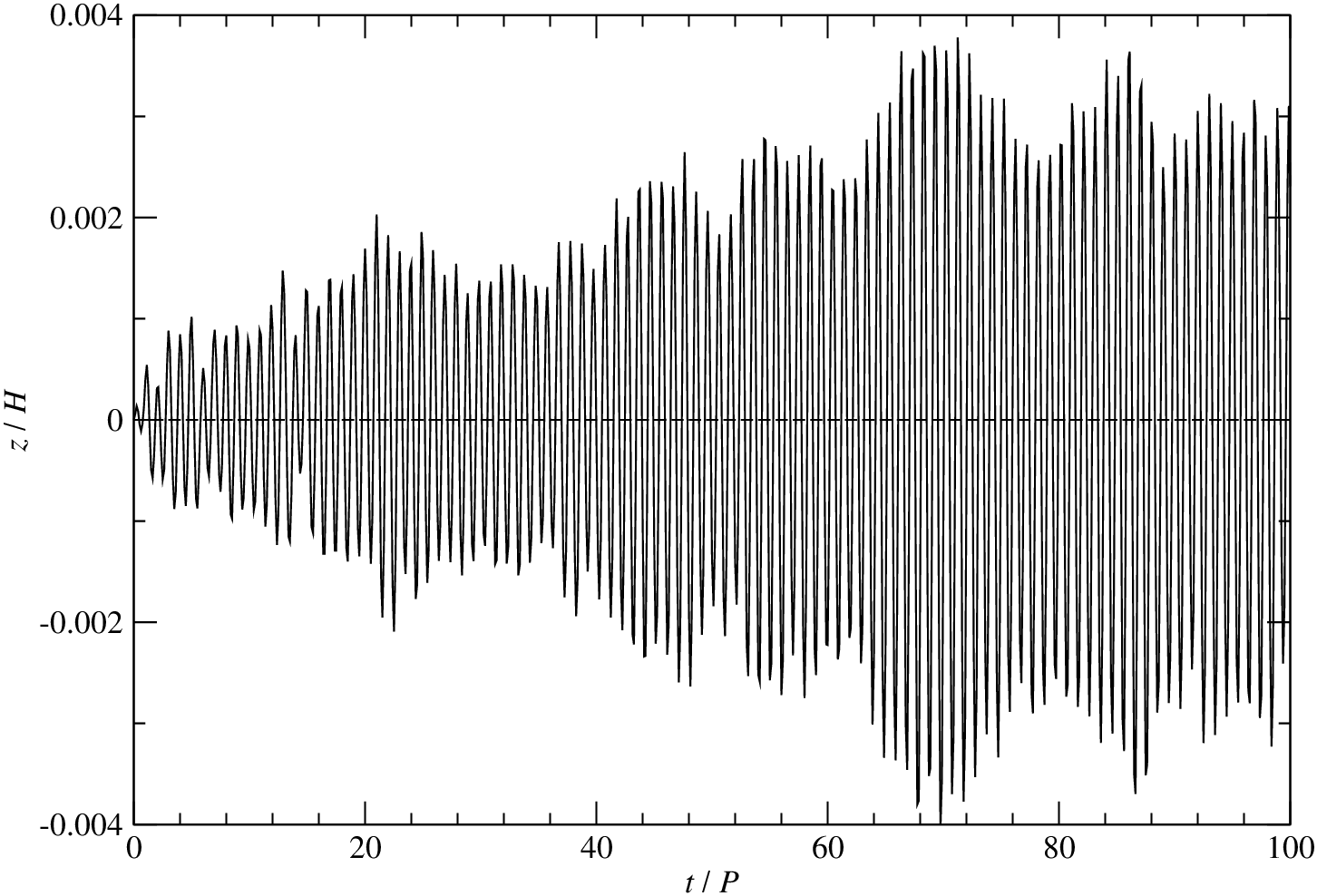}
\caption{Vertical displacement of one example massless particle moving in a stratified, turbulent gas disk.  The simulation box has dimensions of 16$\times$16$\times$4$H$, resolution of 32~points per disk scale height $H$, and strength of disk gravity $\xi = 1$.  The particle has zero initial inclination.}
\label{F:zpt1}
\end{center}
\end{figure}

The only orbital property of a particle that cannot be measured in an unstratified disk model is the inclination $i$.  In a stratified disk, vertical linear gravity from the host star (the third term in the parentheses on the right-hand side of Equation~\eqref{E:eomp2}) provides a restoring force such that particles oscillate in the $z$-direction about the mid-plane, as demonstrated in Figure~\ref{F:zpt1}.  Note that because the particles also experience the gravity of the gas, the period of the oscillation is $P\left(1 + \xi / 4\pi^2\right)^{-1/2}$ in the linear limit (derived from Equations~\eqref{E:eomp2} and \eqref{E:g_0}), which is slightly shorter than the orbital period $P$.  We can calculate the induced inclination (in radians) for a single particle by $i \approx \left(z_\mathrm{max} - z_\mathrm{min}\right) / 2R$, in which $z_\mathrm{max}$ and $z_\mathrm{min}$ are the maximum and the minimum vertical positions in one oscillation, respectively, provided that $z / R \ll 1$.
\placefigure{F:zpt1}

\begin{figure}[htbp]
\begin{center}
\epsscale{.65}
\plotone{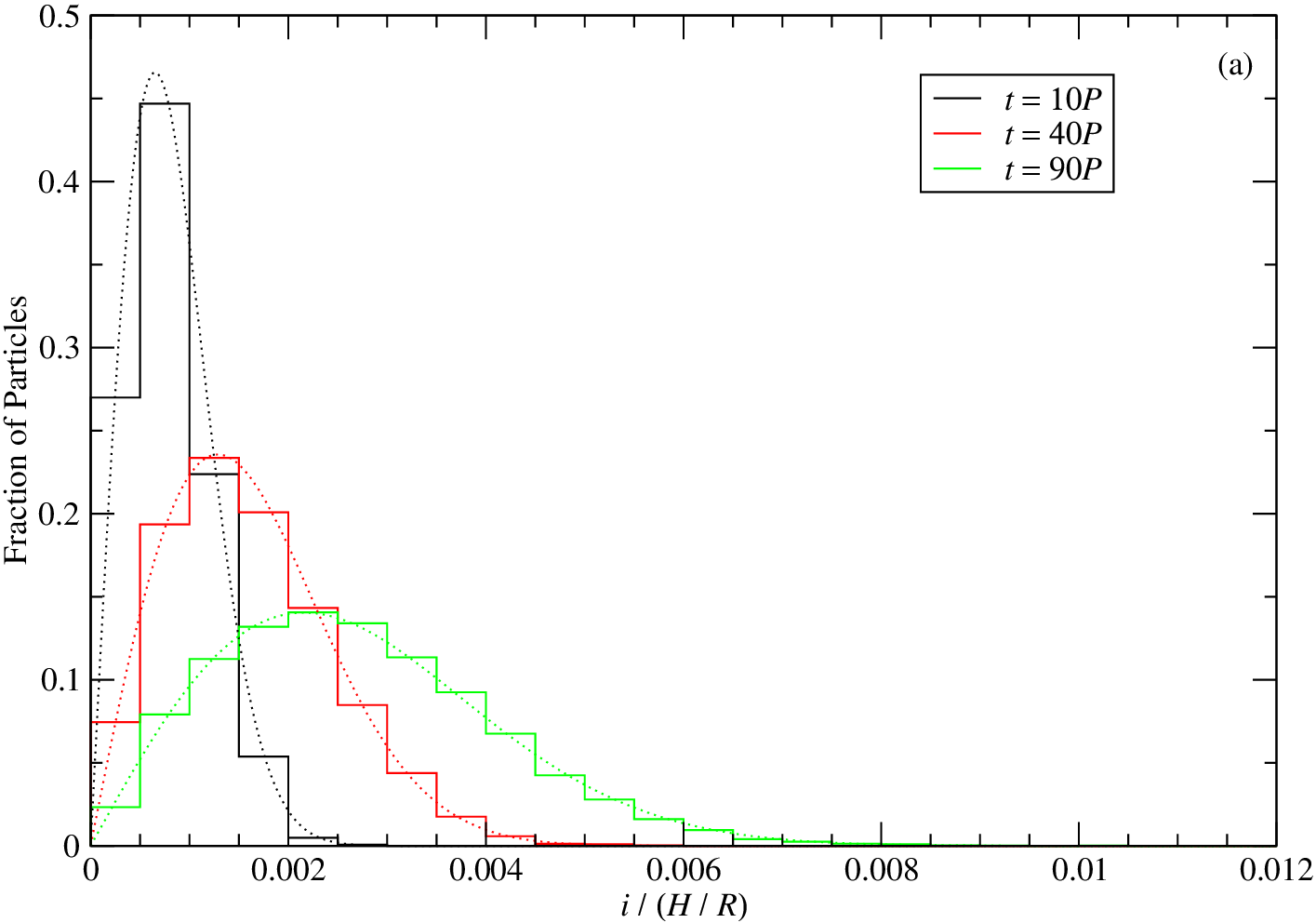}\\[36pt]
\plotone{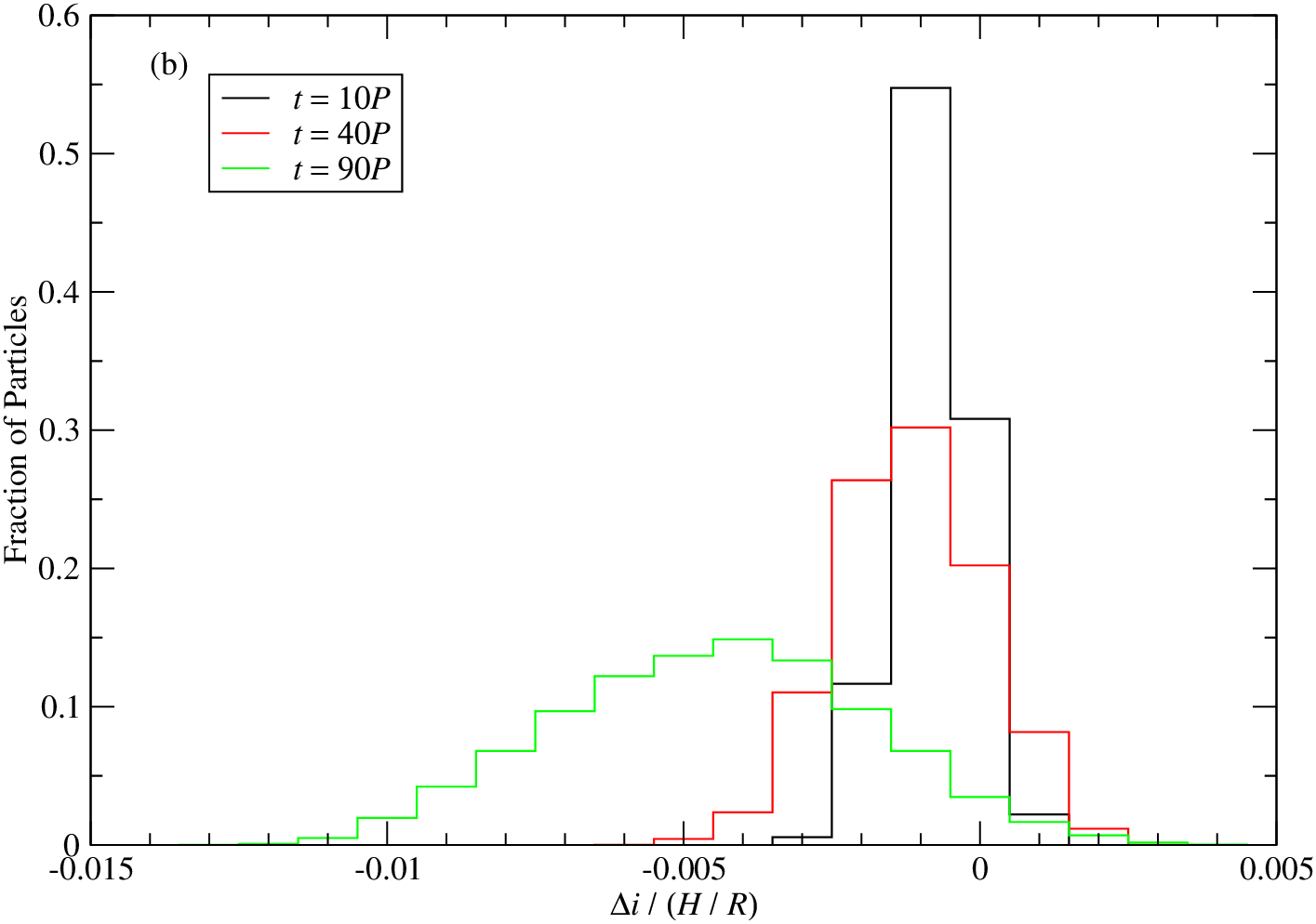}
\caption{Histograms of particles at three different times in bins of (a)~orbital inclination $i$ when the particles have zero initial inclination and (b)~orbital inclination deviation $\Delta i \equiv i - i_0$ when the particles have an initial inclination of $i_0 = 0.1(H/R)$.  The simulation box has dimensions of 8$\times$8$\times$4$H$, resolution of 32~points/$H$, and strength of disk gravity $\xi = 1$.  The dotted-lines in (a) are the best-fits using a Rayleigh distribution, Equation~\eqref{E:dist_i}.}
\label{F:inc}
\epsscale{1}
\end{center}
\end{figure}

Figure~\ref{F:inc}(a) shows the histograms of particles with zero initial inclination sorted in bins of instantaneous inclination at three different times.  Similar to the eccentricity distribution for particles with $e_0 = 0$ described by Equation~\eqref{E:dist_e}, the inclination distribution resembles a time-dependent Rayleigh distribution
\begin{equation} \label{E:dist_i}
f(i,t) = \frac{i}{\sigma_i^2(t)}\exp\left[-\frac{i^2}{2\sigma_i^2(t)}\right],
\end{equation}
where $\sigma_i(t)$ is a time-dependent parameter denoting the width of the distribution.
\placefigure{F:inc}

One might expect the inclination distribution of particles with nonzero initial inclination would evolve as a time-dependent normal distribution with fixed center and increasing width, similar to the eccentricity distribution of particles with nonzero initial eccentricity.  However, Figure~\ref{F:inc}(b) shows different behavior: the distribution increasingly deviates from a normal distribution with time.  The peak leans toward the mid-plane and asymmetry develops with more particles on the left side of the peak (i.e., smaller inclination).  This indicates the diffusion is stronger near the mid-plane, which in fact is consistent with vertical stratification of the gas density.  It would be interesting to measure the dependence of the diffusion coefficient on vertical height, but given our limited sample of initial inclinations, this remains to be investigated.  We therefore focus our attention on the case of particles with zero initial inclination.

\begin{figure}[htbp]
\begin{center}
\plotone{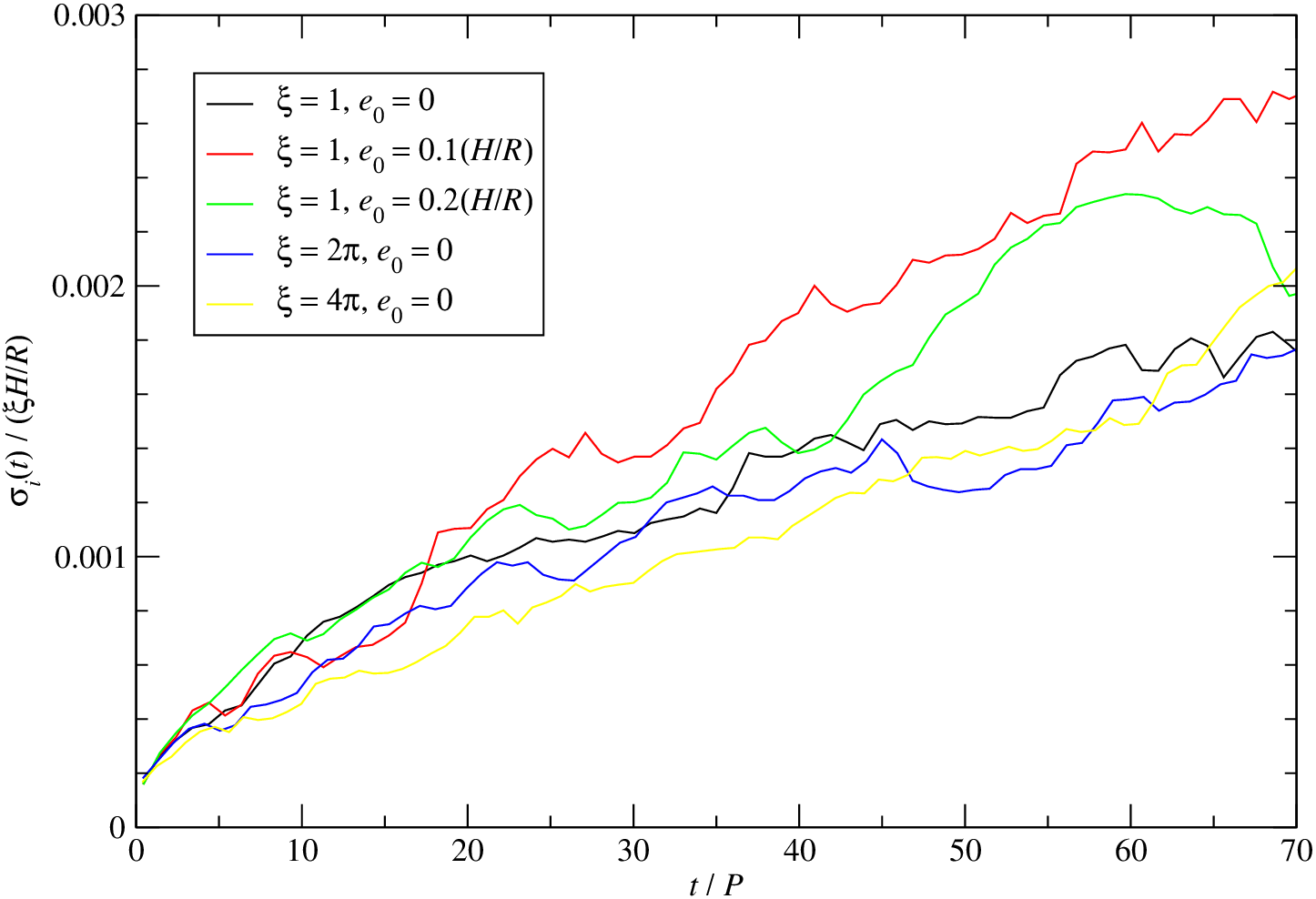}
\caption{Parameter $\sigma_i(t)$ normalized by $\xi H / R$ as a function of time $t$, indicating the evolution of the width of the inclination distribution for disks with varying strength of gravity and particles with varying initial eccentricity.  The simulation box has dimensions of 8$\times$8$\times$4$H$ and resolution of 32~points/$H$.}
\label{F:sigma_i}
\end{center}
\end{figure}

Figure~\ref{F:sigma_i} shows $\sigma_i(t)$ for disks with varying gravity parameter $\xi$ and particles with varying initial eccentricity $e_0$ but zero initial inclination $i_0 = 0$.  Note that when $\sigma_i(t)$ is normalized by $\xi H / R$, all the curves roughly coincide.  This indicates that $\sigma_i(t)$ is linearly dependent on both $\xi H / R$, but independent of $e_0$, similar to what we reported for the evolution of the mean radius and the eccentricity.  The results can thus be summarized by the following expression:
\begin{equation} \label{E:sigma_i}
\sigma_i(t) = C_i\xi\left(\frac{H}{R}\right)\left(\frac{t}{P}\right)^{1/2},
\end{equation}
where $C_i$ is a dimensionless proportionality constant.
Table~\ref{T:orbit} lists our measured values of the constant $C_i$ for different resolutions and box dimensions when particles have zero initial inclination.
\placefigure{F:sigma_i}

We note that in contrast to the orbital radius and the eccentricity, the evolution of the orbital inclination is not significantly affected by the horizontal box size.  The inclination of the particle orbits is mainly affected by the vertical structures of the density field, which in turn is related to the vertical scale height of the density stratification.  The horizontal structures in the magneto-rotational turbulence may have little correlation with the vertical structures so that the horizontal box size has little effect on the vertical motions of the particles.

%-------------------------------------------------------------------------------
\subsection{Velocity Dispersion}

Finally, all three components of the velocity dispersion $\vec{\sigma}_u$ among the particles as a function of time can be measured in our stratified disk models.  Similarly to what we reported in YMM09, each component assumes the same form:
\begin{equation} \label{E:sigma_u}
\sigma_{u,i}(t) = S_i\xi c_s \left(\frac{t}{P}\right)^{1/2},
\end{equation}
where the index $i$ is either $x$, $y$, or $z$ and $S_i$ is the corresponding dimensionless proportionality constant.  We emphasize that $S_y \sim S_x / 2$ always holds because of the fixed ratio of amplitudes in $x$ and $y$ directions in the epicycle motions of the particles, which has been verified in our simulations.  The values of $S_x$ and $S_z$ at different resolutions and box dimensions for the case of particles with zero initial inclination are listed in Table~\ref{T:orbit}.  Note that the value of $S_x$ for the 2$\times$2$\times$4$H$ box at a resolution of 64~points/$H$ is consistent with that from the unstratified disk with the same horizontal size and resolution we reported previously (see Equation~(18) of YMM09), a further confirmation of the consistency between stratified and unstratified models discussed in Sections~\ref{SS:radius} and~\ref{SS:ecc}.

\begin{figure}[htbp]
\begin{center}
\plotone{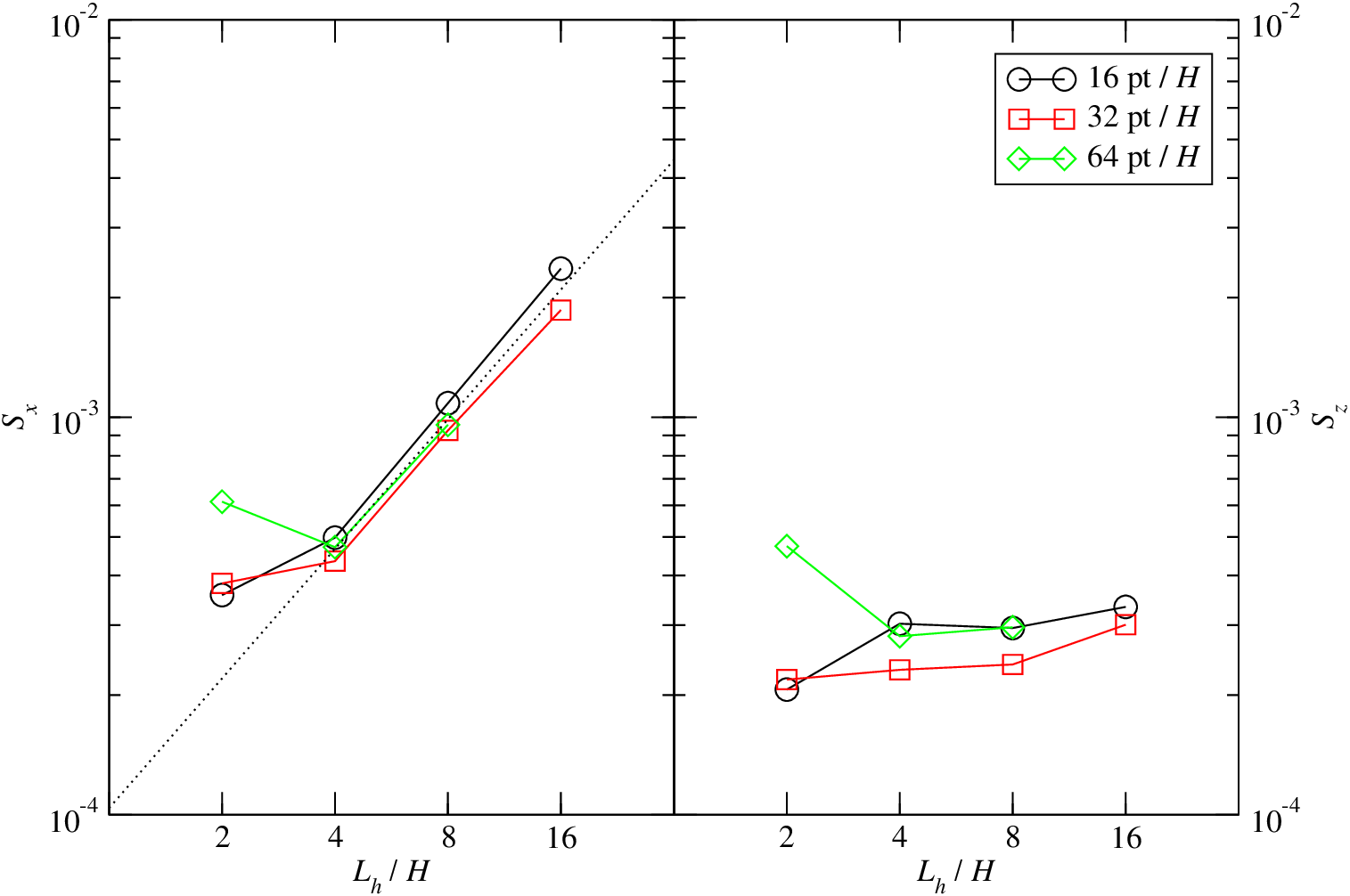}
\caption{Dimensionless proportionality constants $S_x$ and $S_z$ as a function of horizontal box size $L_h$.  They indicate the strength of turbulent excitations for the vertical and radial velocity dispersions of massless particles and are defined in Equation~\eqref{E:sigma_u}.  The dotted line is the power-law fit to $S_x$ for $L_h \geq 4H$.}
\label{F:vel_disp}
\end{center}
\end{figure}

As shown by Figure~\ref{F:vel_disp}, we do not see evident dependence of velocity dispersion on resolution of our models.  On the other hand, the horizontal component $S_x$ (and thus $S_y$) significantly depends on the horizontal box size $L_h$ while the vertical component $S_z$ does not.  This is in accordance with the dependence of the three orbital properties found in previous subsections.  From our measured values for $L_h \geq 4H$, we quantify $S_x$ and $S_z$ with the following expressions:
\begin{eqnarray}
S_x &\simeq& (1.0\times10^{-4})(L_h / H)^{1.08},\label{E:sx}\\
S_z &\simeq& 2.8\times10^{-4}.\label{E:sz}
\end{eqnarray}
\placefigure{F:vel_disp}

%-------------------------------------------------------------------------------
\subsection{Boundary Condition Effects} \label{SS:bce}

\begin{figure}[htbp]
\begin{center}
\epsscale{.7}
\plotone{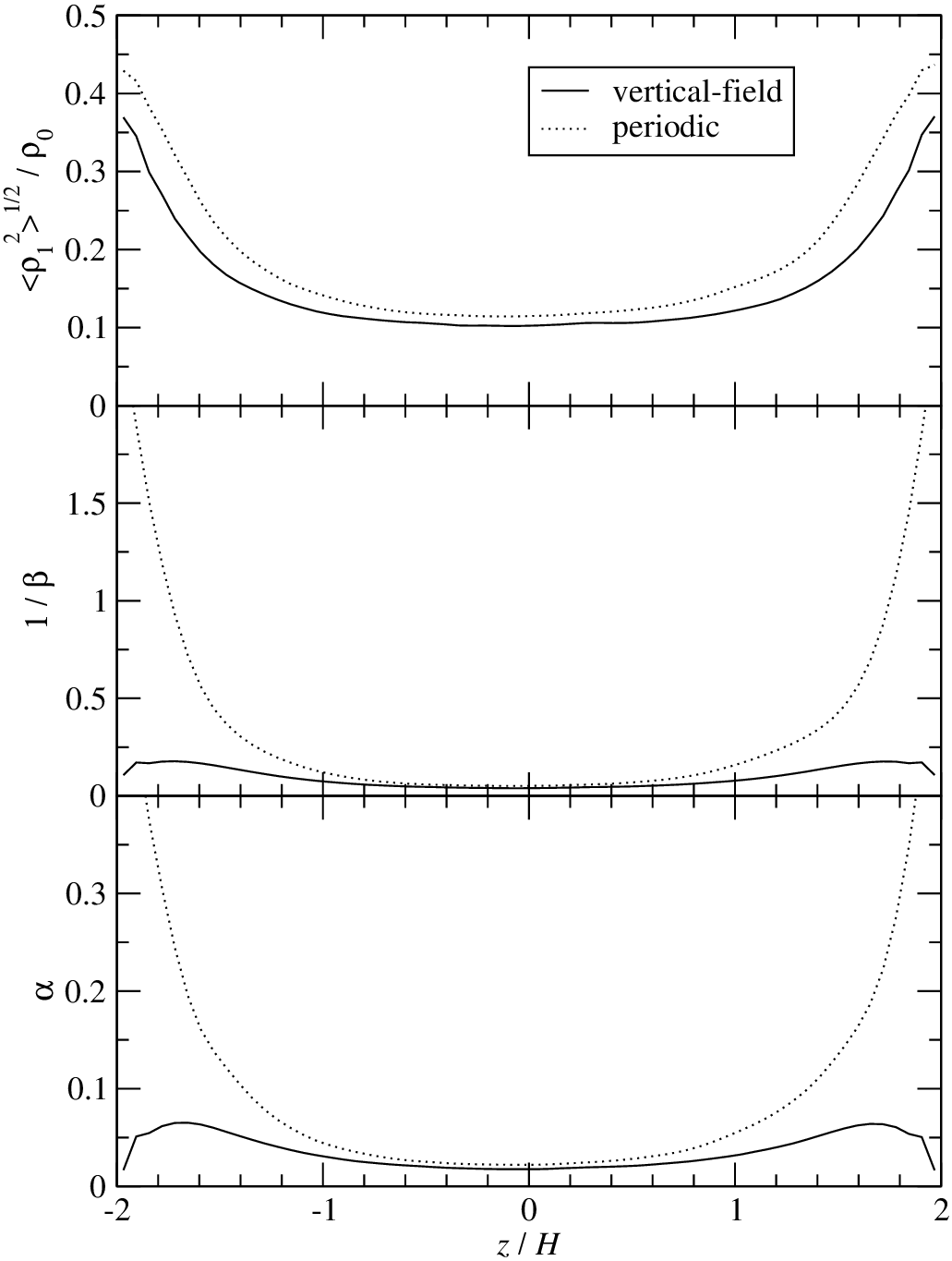}
\caption{Vertical profiles of the magneto-rotational turbulence at saturation stage for $16\times16\times4H$, 16~points/$H$ models with different vertical boundary conditions for the magnetic fields.  The gas density and velocity for both models are periodic.  The panels from top to bottom are rms density fluctuations, inverse plasma $\beta$, and $\alpha$ parameter, respectively.  These horizontally averaged properties are also time averaged over 20~orbital periods, a factor of five shorter than those shown in Figure~\ref{F:mri_z}.}
\label{F:mri_bcz}
\epsscale{1}
\end{center}
\end{figure}

\begin{figure}[htbp]
\begin{center}
\epsscale{.7}
\plotone{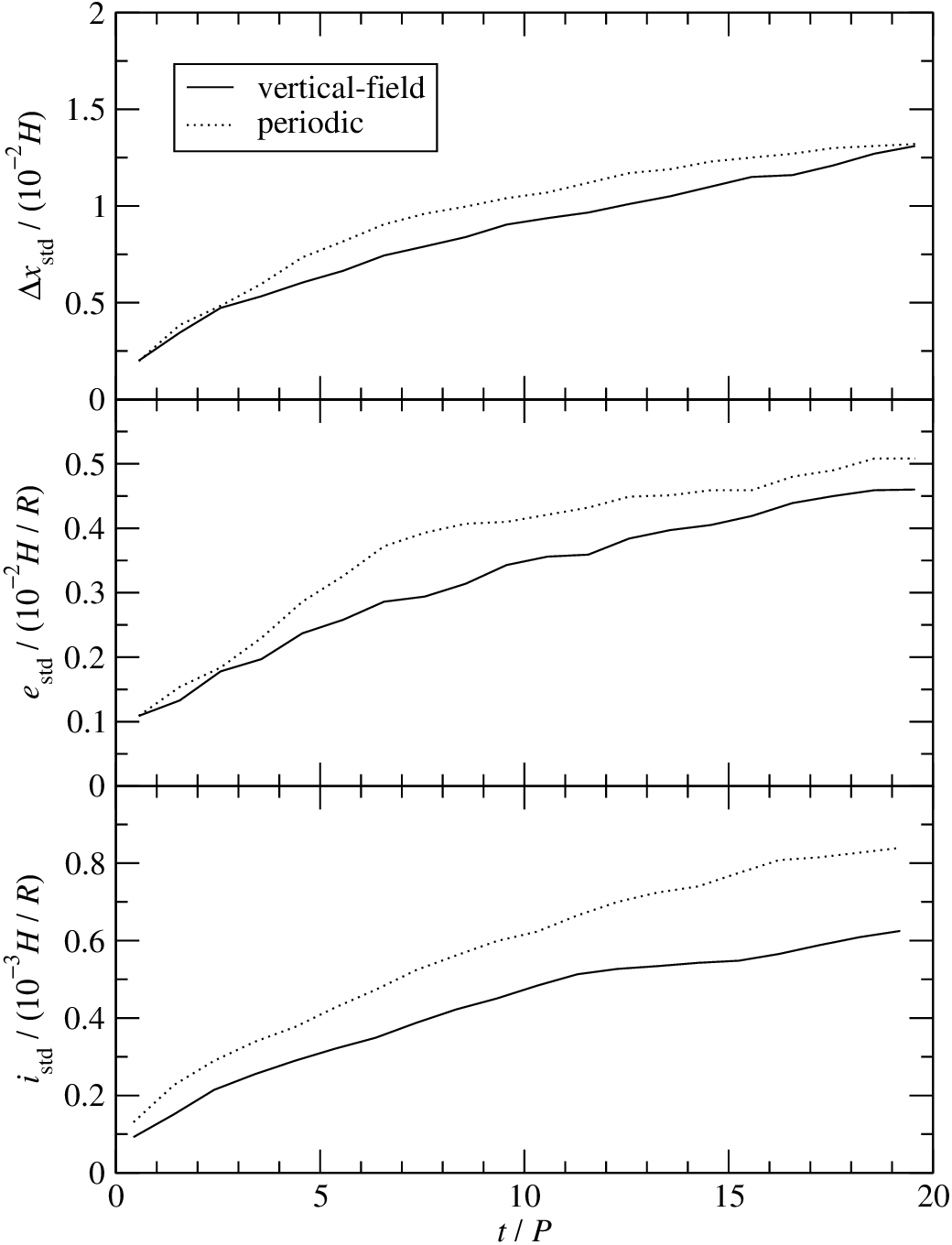}
\caption{Evolution of the particle orbital properties for $16\times16\times4H$, 16~points/$H$ models with different vertical boundary conditions for the magnetic fields.  The gas density and velocity for both models are periodic.  The particles have zero initial eccentricity and inclination.  The panels from top to bottom indicate the standard deviations of the particle distributions in orbital radius, eccentricity, and inclination, respectively, as a function of time.}
\label{F:orbit_bcz}
\epsscale{1}
\end{center}
\end{figure}

In order to confirm that our numerically convenient but perhaps physically inconsistent assumption of vertical field and periodic gas boundary conditions in the $z$-direction does not change our results significantly, we compare turbulence  properties and particle dynamics from models with different vertical boundary conditions for the magnetic fields run at a lower resolution of 16 points per disk scale height $H$ in large boxes with size $16 \times 16 \times 4 H$.  The gas density and velocity remain periodic.  Figure~\ref{F:mri_bcz} shows that within about one disk scale height, the vertical structure of the rms density perturbation, the plasma $\beta$, and the $\alpha$ parameter are in quantitatively close agreement between the models.  Note that the model with periodic boundary  conditions creates a coronal region as low as $|z | \simeq 1.8H$, where $\beta \simeq 1$.  More importantly, Figure~\ref{F:orbit_bcz} shows from top to bottom the time evolution in the standard deviations of particle orbital radius, eccentricity and inclination for the two models.  The orbital radius and eccentricity agree very well irrespective of the vertical boundary conditions, and the effect on inclination differs by only $\sim$30\%, likely due to increased activity at high altitudes for the model with periodic magnetic fields.  Thus, we believe that our results on particle dynamics are robust.
\placefigure{F:mri_bcz}
\placefigure{F:orbit_bcz}

%-------------------------------------------------------------------------------
\section{PROPERTIES OF THE STOCHASTIC FORCE EXERTED BY THE GAS} \label{S:force}

We have demonstrated in Section~\ref{S:turbulence} that various properties of the saturated magneto-rotational turbulence in our simulations converge with both resolution and box dimensions.  However, while converging with resolution, the response of massless particles to the gravity of the turbulent gas does not similarly converge with the horizontal box size, as presented in Section~\ref{S:particles}.  This result raises serious questions about the validity of using the local-shearing-box approximation to simulate the dynamics of any particles under gravitational influence of magneto-rotational turbulence.  In order to understand the lack of convergence with box dimension in the orbital evolution of massless particles, we further study the properties of the gravitational force exerted by the turbulent gas.

%-------------------------------------------------------------------------------
\subsection{Magnitude and Fourier Amplitudes of the Force} \label{SS:force}

Table~\ref{T:force} lists the radial and the azimuthal components of the gravitational force exerted by the turbulent gas, $F_x$ and $F_y$, respectively.  Since the force is stochastic, we ensemble average the root mean square of each force component over all particles, and then report it as time average and $1\sigma$ variation scaled by $\xi m_p H P$, where $m_p$ is the mass of a particle.  Note that in our models, particles only follow the gravitational potential generated by the gas and the host star, so the acceleration of each particle is independent of its mass.  This is an advantage in that the measurements reported in this paper can be used to gauge the gravitational influence of magneto-rotational turbulence on solid particles or planetary objects of any mass and thus its significance compared with other relevant interactions.
\placetable{T:force}

As can be seen in Table~\ref{T:force}, both the radial and the azimuthal components of the force generally increase with the horizontal box size.  It is not clear if there exists any convergence in the radial component towards larger box size, since the 16$\times$16$\times$4$H$ box at a resolution of 32~points per $H$ is the only model that contradicts the general increasing trend.  Certainly, the azimuthal component (i.e., the torque) shows no sign of convergence up to the largest box size we have investigated, with a regression power-law index of about 0.6.  \cite{NG10} also reported in their recent local models a slight trend of increasing torque with increasing horizontal box size (see their Figure~8 and the corresponding discussion).

\begin{figure}[htbp]
\begin{center}
\plotone{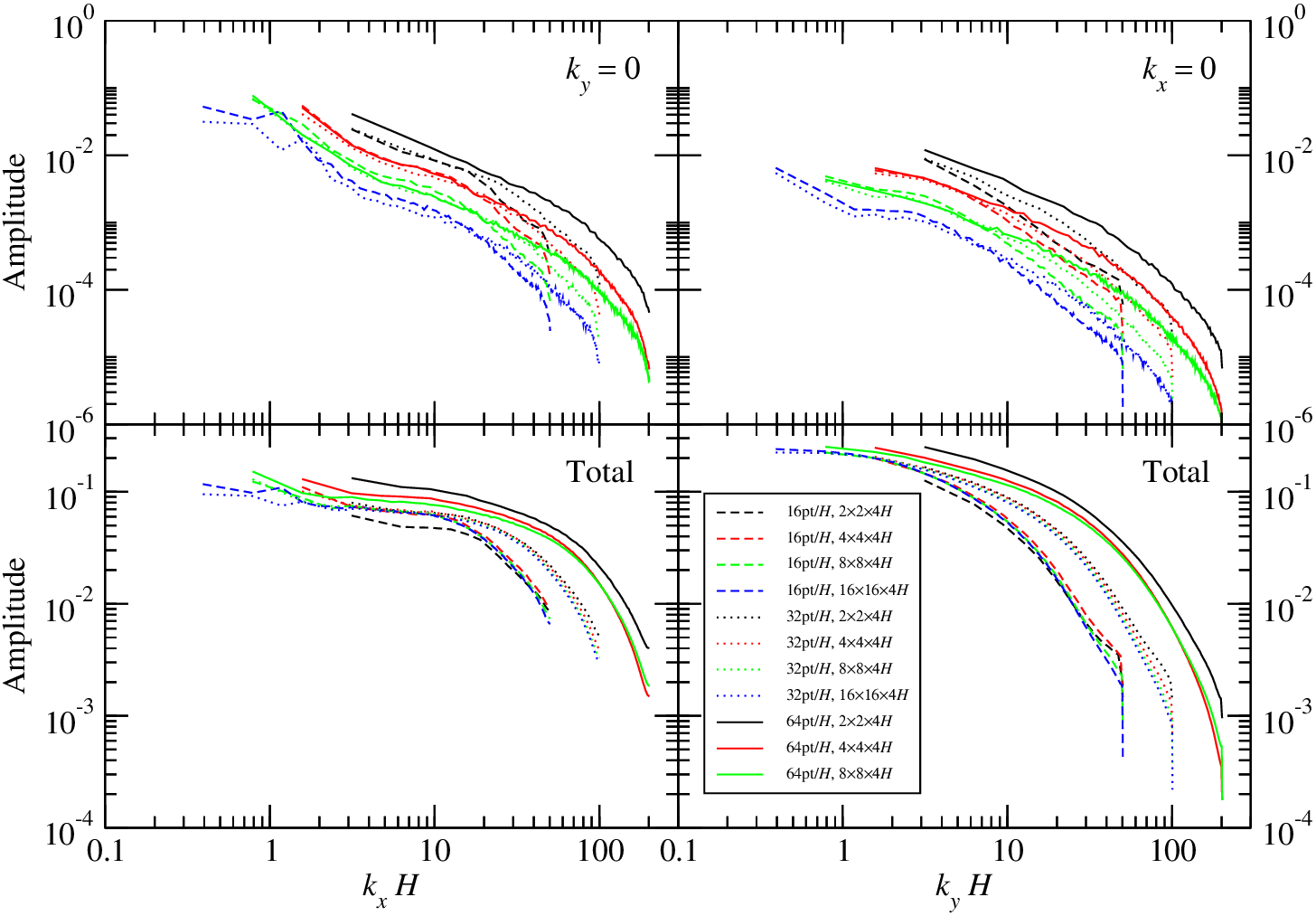}
\caption{Time-averaged Fourier amplitudes of the density fluctuation of saturated magneto-rotational turbulence in the mid-plane for each resolution and box dimensions.  The amplitudes are averaged over a period of over $100P$.  The top-left panel shows $|\tilde{\rho}(k_x,k_y=0,z=0)|$, while the bottom-left panel shows the summation of $|\tilde{\rho}(k_x,k_y,z=0)|$ over $k_y$.  Similarly, the top-right panel shows $|\tilde{\rho}(k_x=0,k_y,z=0)|$, while the bottom-right panel shows the summation of $|\tilde{\rho}(k_x,k_y,z=0)|$ over $k_x$.}
\label{F:fourier}
\end{center}
\end{figure}

Since the density structure of the turbulent gas determines its gravitational force on the particles, it is informative to consider how the structure changes with resolution and horizontal box size.  The density structure of the gas is best analyzed in Fourier space, using the Fourier amplitudes $\tilde{\rho}_1\left(k_x,k_y,z\right)$ defined in Section~\ref{SS:poisson}.\footnote{Note that $\tilde{\rho}_1\left(k_x,k_y,z\right)$ is strictly periodic in sheared coordinates and the wavenumber $\vec{k}$ is slightly different from that of a direct Fourier decomposition of $\rho_1(x,y,z)$; see Section~\ref{SS:poisson}.\label{F:wavenumber}}  Figure~\ref{F:fourier} plots the time-averaged Fourier amplitudes of the gas density in the mid-plane $\tilde{\rho}_1\left(k_x,k_y,z=0\right)$ along either the radial or the azimuthal direction.  The top-left panel shows the amplitude as a function of $k_x$ for $k_y = 0$.  In general, the largest amplitude resides at the longest wavelength while monotonically decreasing with increasing wavenumber \citep{JYK09}.  For any given box dimensions, increasing resolution has little effect on each amplitude, apart from extending the profile toward higher wavenumber.  Flattening of the profile for wavelengths $\gtrsim$8$H$ is hinted for 16$\times$16$\times$4$H$ boxes.  We find no evidence of a downward turn toward even longer wavelength.  On the other hand, increasing the horizontal box size $L_h$ lowers the overall profile of the Fourier amplitudes.  Interestingly, the amplitude of the longest wavelength mode $k_x = k_0 \equiv 2\pi / L_h$ remains roughly constant.
\placefigure{F:fourier}

The bottom-left panel of Figure~\ref{F:fourier} plots the summation of Fourier amplitudes over all $k_y$ at any given $k_x$.  In contrast to the amplitudes for $k_y = 0$, the horizontal box size $L_h$ has little effect on the amplitudes for most of the wavelengths.  Nevertheless, increasing resolution indeed increases the inertial range and resolves more power toward shorter wavelength.

The right-hand column of Figure~\ref{F:fourier} plots the azimuthal counterpart of the left-hand column.  We find similar features in the azimuthal amplitudes as those found in the radial ones.  The only difference is that the flat part of the spectrum in the azimuthal direction is short compared with that in the radial direction, as can be seen in the bottom-right panel.  Power that is not captured in the long-wavelength range of the spectrum in the 2$\times$2$\times$4$H$ boxes is probably the cause for artificially higher amplitude at $(k_x,k_y) = (0,k_0)$ (the top-right panel), which may in turn be responsible for the anomaly found in the particle dynamics shown in Section~\ref{S:particles}.

The Fourier amplitudes shown in Figure~\ref{F:fourier} help us understand the relationship between the gravitational force exerted by the gas and the horizontal box size.
The radial and the azimuthal force components for any given (horizontal) Fourier mode $\vec{k} = (k_x, k_y)$ are related to the corresponding Fourier amplitude $\breve{\rho}(\vec{k})$ by (see Equation~\eqref{E:convol})
\begin{eqnarray}
F_x(\vec{k})
\sim 4\pi G\left(\frac{k_x}{k^2}\right)\left|\breve{\rho}(\vec{k})\right|
\lesssim 4\pi G\frac{\left|\breve{\rho}(k_0,0)\right|}{k_0}
\propto L_h,\label{E:fx}\\
F_y(\vec{k})
\sim 4\pi G\left(\frac{k_y}{k^2}\right)\left|\breve{\rho}(\vec{k})\right|
\lesssim 4\pi G\frac{\left|\breve{\rho}(0,k_0)\right|}{k_0}
\propto L_h,\label{E:fy}
\end{eqnarray}
where we have used the facts that the dominant modes in the radial and the azimuthal directions are the longest wavelength ones $\left|\breve{\rho}(k_0,0)\right|$ and $\left|\breve{\rho}(0,k_0)\right|$, and both are about constant irrespective of resolution and box size.  Therefore, both components should be at most linearly proportional to the horizontal box size in our simulations.  Although involving some simplifications, Equations~\eqref{E:fx} and~\eqref{E:fy} at least qualitatively explain the general trend of increasing stochastic force with horizontal box size.

%-------------------------------------------------------------------------------
\subsection{Correlation Time}

We next evaluate the correlation time $\tau_c$ of the stochastic torque.  This quantity is in fact not well defined in the literature, and different authors have different preferences for its evaluation.  For the purpose of consistency with the framework of radial diffusive migration of particles driven by random torque, we adopt the definition of \citet{JGM06}:
\begin{equation} \label{E:ct1}
  \tau_c \equiv D(J) / \overline{\delta\Gamma^2},
\end{equation}
where $D(J)$ is the radial diffusion coefficient as a function of the Keplerian orbital angular momentum $J$, $\delta\Gamma$ is the fluctuating part of the torque, and the overline denotes an ensemble average.  We repeat the relationship between $D(J)$ and the dimensionless parameter $\epsilon$ here \citep[see][and YMM09]{JGM06}:
\begin{equation} \label{E:diff_coeff}
D(J) = \frac{2.1\times10^{-3}}{16\pi}\epsilon\xi^2
       \left(\frac{H}{R}\right)^2
       \left(\frac{J^2}{P}\right).
\end{equation}

To calculate the correlation time $\tau_c$ with our definition, both the diffusion coefficient $D(J)$ and the magnitude of the stochastic torque are needed.  As emphasized in YMM09, we have all the information about the particle distribution against each orbital property as a function of time, and thus we can directly evaluate the diffusion coefficient in the context of a diffusion equation, which is reported in Section~\ref{SS:radius} and Table~\ref{T:orbit}.  Furthermore, we also have the ensemble-averaged rms value of the stochastic torque, which is reported in Section~\ref{SS:force} and Table~\ref{T:force}.  Equations~\eqref{E:ct1} and~\eqref{E:diff_coeff} can then be used to calculate the correlation time $\tau_c$, and the result is listed in Table~\ref{T:force}.

As indicated in Table~\ref{T:force}, the correlation time of the stochastic torque $\tau_c$ does not appreciably depend on the resolution but it is significantly affected by the horizontal box size.  The larger the horizontal box size, the longer the correlation time of the torque; this trend was also reported by \citet{NG10}.  Up to the largest box size we have investigated ($L_h \lesssim 16H$), it is not clear if $\tau_c$ will converge with $L_h$.  This behavior is similar to that of the torque magnitude, as discussed in Section~\ref{SS:force}.

As a side note, one may want to evaluate the correlation time of the stochastic torque $\tau_c$ without referring to stochastic particle orbital evolution and invert Equation~\eqref{E:ct1} to estimate the radial diffusion coefficient $D(J)$ indirectly.  We have shown in YMM09, for the case of ideal unstratified disks, that the following formula, motivated by the definition of $D(J)$ under the framework of the Fokker-Planck formalism, serves this purpose:
\begin{equation} \label{E:ct2}
  \tau_c \approx
  \frac{\int_0^{\infty} \overline{\mathrm{ACF}(\tau)}\,\mathrm{d}\tau}
       {2\overline{\mathrm{ACF}(0)}},
\end{equation}
where $\mathrm{ACF}(\tau)$ is the autocorrelation function of the stochastic torque as a function of the time lag $\tau$.
Here we repeat the exercise and use Equation~\eqref{E:ct2} to estimate $\tau_c$ for the case of ideal stratified disks.  The result is listed in Table~\ref{T:force} and shows remarkable consistency with the values obtained directly from the evolution of particle orbital distribution.  Therefore, Equation~\eqref{E:ct2} remains a useful estimator of the correlation time $\tau_c$.

\begin{figure}[htbp]
\begin{center}
\plotone{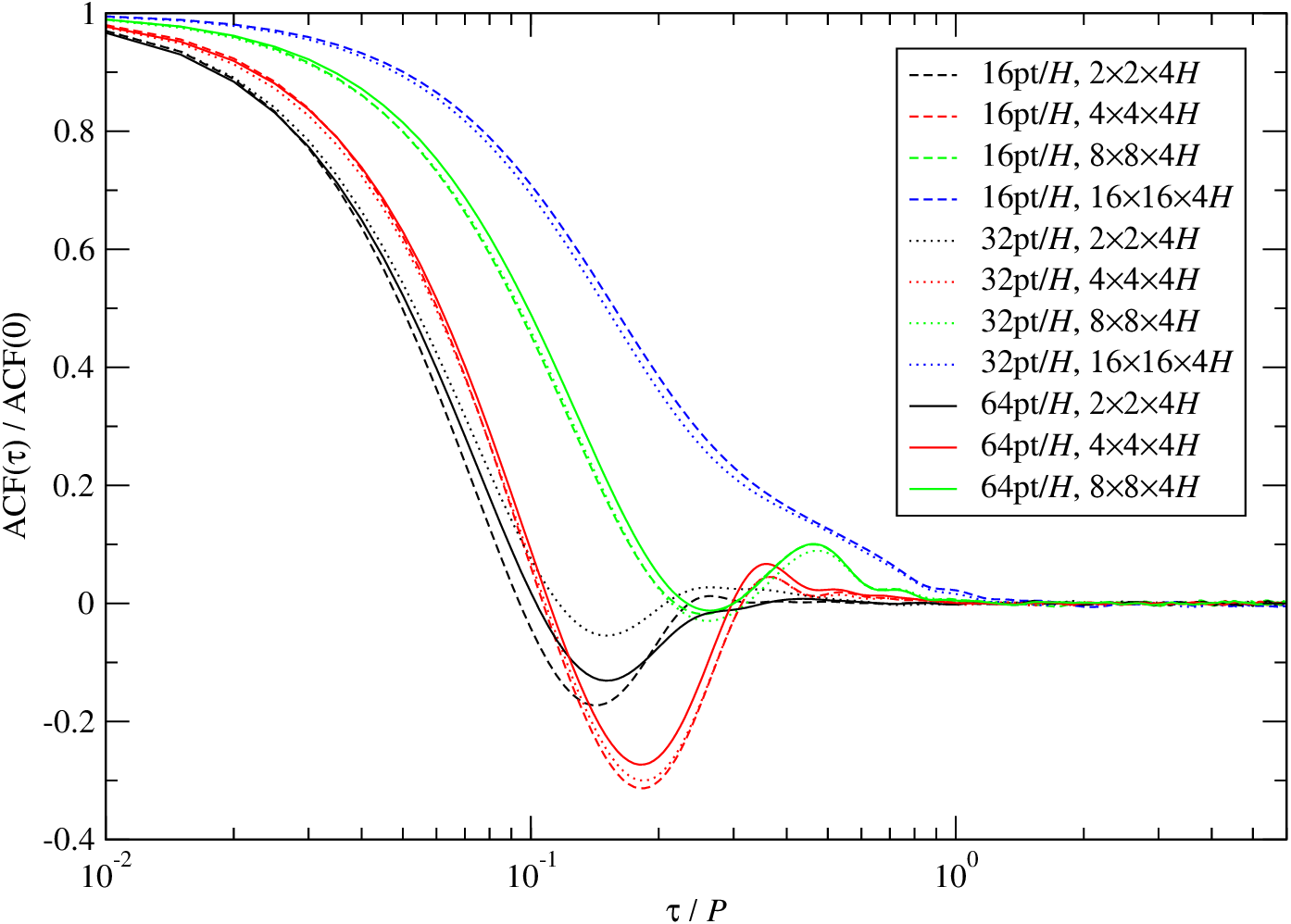}
\caption{Normalized, ensemble-averaged autocorrelation function of the stochastic torque exerted by the turbulent gas for each resolution and box dimensions.}
\label{F:acf}
\end{center}
\end{figure}

To further understand the trend of increasing correlation time with increasing horizontal box size, we plot the autocorrelation function of the stochastic torque in Figure~\ref{F:acf}.  As also demonstrated by \citet{NG10}, the larger the box, the less oscillatory the autocorrelation function is and the less pronounced is the negative contribution to the radial diffusion of the particles.  This behavior in turn is consistent with increasing values of the correlation time and the radial diffusion coefficient.
\placefigure{F:acf}

In addition to the stochastic torque, a similar analysis may be conducted for the radial component of the gravitational force exerted by the gas.  We find that the correlation time of the radial force also increases with the horizontal box size.  However, our models indicate a much longer correlation timescale, which may amounts to several tens of orbital periods.  Given that our simulations only lasted for about 100$P$, the magnitude of the correlation time is not well calculated and much longer simulations may be required to cover the full spectrum of the autocorrelation function for the radial force.

\citet{JYK09} have discussed the dominant role of the longest wavelength, purely radial Fourier mode $\vec{k} = (k_0,0)$ in great detail and suggested that these kind of persistent structures are similar to the zonal flows found in many other astrophysical systems.  As noted by the authors, these axisymmetric structures can not generate torques that affect the orbital radius of the embedded particles.  While the radial forces due to the zonal flows can still affect the eccentricity of the particles, their long correlation time indicates that they may not be responsible for the diffusive evolution of the particle eccentricity, as shown in YMM09 and Section~\ref{SS:ecc}.\footnote{It remains possible that the stochastic radial forces due to small-scale axisymmetric density perturbations drive the diffusive evolution of the particle eccentricities.}  On the other hand, the longest wavelength, purely azimuthal Fourier mode  $\vec{k} = (0, k_0)$ in the gas structure\footnote{Note that this mode does not represent a single shearing wave, and the wavenumber $(k_x,k_y)$ is independent of time; see footnote~\ref{F:wavenumber} and Section~\ref{SS:poisson}.} is significant enough that both the radial diffusive migration and the eccentricity evolution of the particles are affected by the large-scale azimuthal density perturbations and thus the box dimensions.

%-------------------------------------------------------------------------------
\section{RELATION OF BOX SIZE TO PARTICLE ORBITAL EVOLUTION} \label{S:box_size}

As described in Section~\ref{S:force}, the magnitude and the correlation time of the gravitational force exerted by the turbulent gas correlates with the horizontal size of our local shearing box.  This behavior essentially explains the correlation of the stochastic evolution of the particle orbital radius and eccentricity with the horizontal box size reported in Section~\ref{S:particles}.  These results challenge the validity of the local shearing box to describe large-scale density structures in magneto-rotational turbulence.

By measuring the two-point correlation functions of density, velocity, and magnetic field in magneto-rotational turbulence, \citet{GG09} found extended density features in contrast to well localized magnetic structures.  The authors suggested that the density features are propagating acoustic waves excited by the turbulence \citep[see also][]{HP09a,HP09b}.  It is not clear yet what the dissipation length scale for these waves is, which is crucial to understanding whether a local shearing box can capture the largest-scale structures in the density fluctuations excited by magneto-rotational turbulence.

On the other hand, a global disk model may require high resolution in order to self-consistently produce large-scale structures.  For example, \citet{JYK09} argued that the inverse cascade of magnetic energy from small scales to large scales might be responsible for ultimately launching zonal flows.  To confirm this mechanism in a global context, a model that is capable of resolving at least the correlation lengths in the magnetic structures might be necessary.

%-------------------------------------------------------------------------------
\subsection{Bridge to a Global Disk Model?}

Given the correlation with horizontal box size reported in this work, it is not clear how a local shearing box can be connected to a global disk model so that both models give consistent results on the particle dynamics under the gravitational influence of density fluctuations in magneto-rotational turbulence.

Nevertheless, we conjecture at this point that the criterion $L_h / R \sim O(1)$ might provide a physical length scale for a local shearing box.  At this scale, the local-shearing-box approximation formally breaks down since it assumes $L / R \ll 1$.  The curvature terms become important and this might trigger turbulent eddies to damp propagating waves.  If this conjecture could be verified, Equations~\eqref{E:cx}, \eqref{E:ce} and~\eqref{E:sx} would prove useful in evaluating the gravitational influence of the turbulent gas on particle dynamics.  Substituting $R$ for $L_h$ in these equations, our models predict that
\begin{equation} \label{E:cs_predict}
  \left\{
  \begin{array}{ccl}
  C_x &\simeq& 6.6\times10^{-5}\,(H/R)^{-1.35}\\
  C_e &\simeq& 7.2\times10^{-5}\,(H/R)^{-1.08}\\
  S_x &\simeq& 1.0\times10^{-4}\,(H/R)^{-1.08}
  \end{array}
  \right.\qquad \textrm{for $\alpha \sim 10^{-2}$.}
\end{equation}

\citet{NG10} presented the most recent work on stochastic planetesimal orbital evolution in a global disk model.  Their measurement of the orbital radius and radial velocity dispersion can be directly compared with ours.  These authors reported in their global model G3, whose $\alpha\simeq0.02$ was the closest to ours, that $\sigma_x(t) = 7.2\times10^{-4}\,R\sqrt{t/P}$ and $\sigma_{u,x}(t) = 8.2\times10^{-3}\,c_s\sqrt{t/P}$.  In this specific model, a disk aspect ratio\footnote{See footnote~\ref{scale_height}.} of $H / R \approx 0.071$ and a strength of gas gravity of $\xi \approx 2.1$ at 5~AU were adopted.  In comparison, Equations~\eqref{E:sigma_x}, \eqref{E:sigma_u}, and~\eqref{E:cs_predict} give $\sigma_x(t) \approx 6.9\times10^{-3}\,(H/\sqrt{2})\sqrt{t/P} = 3.4\times10^{-4}\,R\sqrt{t/P}$ and $\sigma_{u,x}(t) \approx 3.8\times10^{-3}\,c_s\sqrt{t/P}$.  Our results are only about a factor of two lower than those of \citet{NG10}.  Therefore, the criterion $L_h \sim R$ for a local shearing box seems to offer some consistency between global and local models.

These comparisons are not meant to be accurate and conclusive, though, since the proposed criterion $L_h / R \sim O(1)$ invites significant uncertainty.  Further investigation into the large-scale structures of magneto-rotational turbulence and the discrepancy between global and local disk models is still needed.

%-------------------------------------------------------------------------------
\subsection{Implications for Planet Formation and Migration}

If the criterion $L_h \sim R$ holds, Equations~\eqref{E:epsilon} and~\eqref{E:gamma} imply
\begin{equation} \label{E:eg_predict}
  \left\{
  \begin{array}{ccl}
  \epsilon &\simeq& 6.5\times10^{-6}\,(H / R)^{-2.69}\\
  \gamma &\simeq& 3.6\times10^{-5}\,(H / R)^{-1.08}
  \end{array}
  \right.\qquad\textrm{for}\ \alpha \sim 10^{-2}.
\end{equation}
Consider a region at $R = 5$~AU in a minimum-mass solar nebula \citep[MMSN;][]{cH81}, where $H / R \approx 0.08$.  Equations~\eqref{E:eg_predict} give $\epsilon \approx 7\times10^{-3}$ and $\gamma \approx 6\times10^{-4}$.  This value of the $\epsilon$ parameter is about one order of magnitude higher than was reported in YMM09 for the case of a 2$\times$2$\times$2H unstratified box, while the value of the $\gamma$ parameter is only about a factor of three larger.  Note that according to Equations~\eqref{E:eg_predict}, both $\epsilon$ and $\gamma$ decrease with increasing disk aspect ratio $H / R$, which often increases with increasing radial distance $R$.

We suggested in YMM09 that in a typical protoplanetary disk, the radial diffusive migration of protoplanets induced by magneto-rotational turbulence may be unimportant compared to secular migration.  According to \citet{JGM06}, for the diffusive migration to be able to dominate over type I migration, the $\epsilon$ parameter should be greater than or about 0.1--1 for the MMSN.  Our new measurement remains orders of magnitude smaller than this transition value.  Therefore, our previous conclusions on the insignificance of planetary diffusive migration may still hold even though $\epsilon$ is increased by an order of magnitude.  However, note that recent calculations for type I migration in non-isothermal disks \citep[e.g.,][]{LPM10,PB10,PB11} may negate the dominance of secular migration over stochastic migration driven by magneto-rotational turbulence.

We also argued in YMM09 that kilometer-sized planetesimals moving in magneto-rotational turbulence survive mutual collisional destruction, except in the inner region of a young protoplanetary disk.  This was based on the results of \citet{IGM08} for the cases of $\gamma = 10^{-3}$ and $10^{-4}$.  Since our new measurement does not fall outside this range,  the same conclusion still applies.

In contrast, \citet{NG10} argued that the velocity dispersion of kilometer-sized planetesimals excited by magneto-rotational turbulence might be so large that these planetesimals should be erosive to each other.  Before a consistent scenario for the survivability of planetesimals could be assembled, however, several factors remain to be accounted for.  First, eccentricity damping due to tidal interaction between a planetesimal and its surrounding gas acts to circularize the orbits of the planetesimals \citep[e.g.,][]{GT80}.  This mechanism was absent in the models of \citet{NG10} and thus their measurement of the velocity dispersion of planetesimals should be considered as an upper limit.\footnote{This is not true if mutual gravitational scattering between planetesimals dominates over turbulence-driven excitation.  However, this might not be the case in a typical protoplanetary environment; see YMM09.}  On the other hand, using analytical arguments to estimate the equilibrated eccentricity of planetesimals, \citet{IGM08} took three possible eccentricity damping effects into account --- tidal interaction, gas drag, and inelastic collisions --- and thus their estimate of the velocity dispersion of planetesimals might be more realistic.  Even though significant uncertainty was involved in the assumed strength of turbulent excitation of eccentricity in the models of \citet{IGM08}, it could be improved by more accurate measurement of the $\gamma$ parameter, as provided by the present work.

Other interesting progress in the study of mutual collisions of planetary objects includes different outcomes from different impact angles \citep[head-on vs.\ hit-and-run scenarios; see, e.g.,][]{eA09,MS09} and improved calculations on material properties \citep[e.g.,][]{LS09,SL09,jW10}.  Given all these new developments, a reconsideration of the survivability of kilometer-sized planetesimals moving in magneto-rotational turbulence seems warranted.

%-------------------------------------------------------------------------------
\section{SUMMARY} \label{S:summary}

Directly extending our previous publication (YMM09), we continue to study massless particles moving under the gravitational influence of density fluctuations due to saturated magneto-rotational turbulence in a local, isothermal, Keplerian gas disk.  We include linearized vertical gravity from the host star and thus vertical stratification of the gas disk.  For comparison, the conditions in the mid-plane of the vertically stratified disks are exactly the same as those in the unstratified disks of YMM09.

In order to accurately measure the gravitational effect of the turbulent gas, we separate the gas density $\rho$ into two components: the basic state $\rho_0(z)$ for the vertical hydrostatic equilibrium (Equation~\eqref{E:rho_0}) and the density deviation $\rho_1 \equiv \rho - \rho_0$ from this basic state.  We use the exact gravitational acceleration due to the basic state (Equation~\eqref{E:g_0}) and only solve the Poisson equation for the gravitational potential due to the density deviation (Equation~\eqref{E:poisson}).  We emphasize that since the Poisson equation is linear in density, this approach does not assume small density fluctuations.  Furthermore, we implement isolated boundary conditions in the vertical direction and thus any density fluctuation outside the vertical computational domain is neglected.

By imposing a weak, uniform external magnetic field, we maintain a constant level of saturated magneto-rotational turbulence in the disk mid-plane.  Several turbulence properties demonstrate convergence with both resolution up to 64~points per disk scale height $H$ and horizontal box size up to 16$H$.  The \citet{SS73} $\alpha$ parameter in the mid-plane of our models is controlled at the level of $\sim$10$^{-2}$.

However, even though the properties of the turbulent gas appear numerically convergent, the dynamics of massless particles moving under the gravity of this turbulent gas does not converge with the horizontal box size $L_h$.  The larger the horizontal box size, the stronger the gravitational effect of the gas on the particles.  Specifically, the evolution of the orbital radius, the eccentricity, and the horizontal velocity dispersion of the particles is roughly linearly dependent on $L_h$ up to 16$H$.  This trend was also found in our unstratified models (YMM09), and we find consistency between the unstratified models and the mid-plane of the stratified models.  In contrast to the horizontal components of the particle movement, we find that the evolution of the inclination and the vertical velocity dispersion is not significantly affected by $L_h$.

The dependence of particle dynamics on the horizontal box size can be traced back to the density structure of the gas.  Consistent with the large-box models studied by \citet{JYK09}, the longest wavelength Fourier mode dominates the density spectrum along the radial direction in our models.  Furthermore, we find that the longest wavelength Fourier mode in the azimuthal direction also strongly influences the particle dynamics, leading to the diffusive evolution of both the orbital radius and the eccentricity of the particles.  The spectral amplitudes of these longest wavelength modes are roughly constant against the horizontal box size.  Using a simple single-mode analysis, we show that the linear dependence of the particle response is a natural outcome of these findings for the density spectrum.

Correlation of particle dynamics with box size poses a major difficulty for the interpretation of local-shearing-box models involving gravitational physics of magneto-rotational turbulence.  We can nevertheless conjecture that $L_h \sim R$, where $R$ is the distance of the box center to the host star, might be a natural scale of choice for a local model to approach reality.  If this conjecture holds, we find that our previous conclusions in YMM09 on the unimportance of radial diffusive migration for protoplanets as well as the survivability of kilometer-sized planetesimals under collisional destruction may still be valid.  Ultimately, high-resolution global disk models and detailed comparisons with large-box local models might be necessary to settle this issue.

\acknowledgments
We thank the anonymous referee for critical comments that significantly improved the rigor and clarity of this paper, and thank Charles Gammie, Anders Johansen, Roman Rafikov, Cl\'{e}ment Baruteau, Martin Pessah, Richard Nelson, and Oliver Gressel for their insightful discussion on this research.  This research was supported in part by the Perimeter Institute for Theoretical Physics.  Resources supporting this work were provided by the NASA High-End Computing (HEC) Program through the NASA Advanced Supercomputing (NAS) Division at Ames Research Center.  Partial support of this work was provided by the NASA Origins of Solar Systems Program under grant NNX07AI74G.

%-------------------------------------------------------------------------------

%-------------------------------------------------------------------------------
\clearpage

\begin{deluxetable}{rcccccc}
\rotate
\tablecolumns{7}
\tablewidth{0in}
\tablecaption{Mid-plane Properties of the Magneto-Rotational Turbulence\label{T:mri}}
\tablehead{ 
\colhead{Dimensions} & \colhead{Resolution}
& \colhead{$\sqrt{\langle\rho_1^2\rangle} / \rho_m$}
& \colhead{$\beta^{-1}$}
& \colhead{$\langle B_x B_y\rangle / \mu_0\rho_m c_s^2$}
& \colhead{$\langle\rho u_x u_y\rangle / \rho_m c_s^2$}
& \colhead{$\alpha$}\\
\colhead{[$H$]}
& \colhead{[pt/$H$]}
& \colhead{[10$^{-1}$]}
& \colhead{[10$^{-2}$]}
& \colhead{[10$^{-2}$]}
& \colhead{[10$^{-2}$]}
& \colhead{[10$^{-2}$]}}
\startdata
2$\times$2$\times$4 & 16 & 0.9$\pm$0.2 & \phantom{0}5.3$\pm$2.6 & 1.9$\pm$0.8
& 0.5$\pm$0.3 & 1.2$\pm$0.4 \\ 
4$\times$4$\times$4 & 16 & 1.4$\pm$0.2 & \phantom{0}5.9$\pm$1.8 & 2.4$\pm$0.7
& 0.7$\pm$0.2 & 1.5$\pm$0.4 \\ 
8$\times$8$\times$4 & 16 & 1.6$\pm$0.3 & \phantom{0}5.2$\pm$0.8 & 2.2$\pm$0.3
& 0.7$\pm$0.1 & 1.4$\pm$0.2 \\ 
16$\times$16$\times$4 & 16 & 1.7$\pm$0.3 & \phantom{0}5.1$\pm$0.4 & 2.2$\pm$0.2
& 0.7$\pm$0.1 & 1.4$\pm$0.1 \\ 
2$\times$2$\times$4 & 32 & 1.0$\pm$0.3 & \phantom{0}7.6$\pm$3.4 & 2.8$\pm$1.0
& 0.6$\pm$0.3 & 1.6$\pm$0.5 \\ 
4$\times$4$\times$4 & 32 & 1.2$\pm$0.3 & \phantom{0}6.3$\pm$1.8 & 2.6$\pm$0.6
& 0.7$\pm$0.2 & 1.6$\pm$0.4 \\ 
8$\times$8$\times$4 & 32 & 1.5$\pm$0.4 & \phantom{0}5.1$\pm$0.5 & 2.3$\pm$0.2
& 0.6$\pm$0.1 & 1.4$\pm$0.1 \\ 
16$\times$16$\times$4 & 32 & 1.3$\pm$0.1 & \phantom{0}5.1$\pm$0.4 & 2.3$\pm$0.1
& 0.6$\pm$0.1 & 1.4$\pm$0.1 \\ 
2$\times$2$\times$4 & 64 & 1.6$\pm$0.6 & 17.1$\pm$9.4 & 5.5$\pm$3.0
& 0.9$\pm$0.6 & 3.0$\pm$1.5 \\ 
4$\times$4$\times$4 & 64 & 1.4$\pm$0.3 & \phantom{0}8.1$\pm$1.6 & 3.4$\pm$0.7
& 0.8$\pm$0.2 & 2.0$\pm$0.4 \\ 
8$\times$8$\times$4 & 64 & 1.6$\pm$0.4 & \phantom{0}7.0$\pm$0.9 & 3.0$\pm$0.4
& 0.6$\pm$0.1 & 1.7$\pm$0.2
\enddata 
\end{deluxetable}

%-------------------------------------------------------------------------------
\clearpage

\begin{deluxetable}{rcccccccc}
\tablecolumns{7}
\tablewidth{0pc}
\rotate
\tablecaption{Dynamical Properties of the Massless Particles with $e_0 = 0$ and $i_0 = 0$\label{T:orbit}} 
\tablehead{
\colhead{Dimensions} & \colhead{Resolution}
& \colhead{$C_x$} & \colhead{$C_e$} & \colhead{$C_i$}
& \colhead{$S_x$} & \colhead{$S_z$}
& \colhead{$\epsilon$} & \colhead{$\gamma$}\\
\colhead{[$H$]} & \colhead{[pt/$H$]} & & & & & & &}
\startdata
2$\times$2$\times$4 & 16 & 2.4(-4) & 2.4(-4) & 2.0(-4) & 3.6(-4) & 2.1(-4)
                         & 8.3(-5) & 1.2(-4)\\ 
4$\times$4$\times$4 & 16 & 4.4(-4) & 3.5(-4) & 2.9(-4) & 5.0(-4) & 3.0(-4)
                         & 2.9(-4) & 1.7(-4)\\ 
8$\times$8$\times$4 & 16 & 1.3(-3) & 7.4(-4) & 2.8(-4) & 1.1(-3) & 2.9(-4)
                         & 2.6(-3) & 3.7(-4)\\ 
16$\times$16$\times$4 & 16 & 3.0(-3) & 1.6(-3) & 2.4(-4) & 2.4(-3) & 3.3(-4)
                           & 1.4(-2) & 8.0(-4)\\ 
2$\times$2$\times$4 & 32 & 2.6(-4) & 2.7(-4) & 2.6(-4) & 3.8(-4) & 2.2(-4)
                         & 1.0(-4) & 1.3(-4)\\ 
4$\times$4$\times$4 & 32 & 3.7(-4) & 3.0(-4) & 2.5(-4) & 4.3(-4) & 2.3(-4)
                         & 2.1(-4) & 1.5(-4)\\ 
8$\times$8$\times$4 & 32 & 1.1(-3) & 6.4(-4) & 1.7(-4) & 9.3(-4) & 2.4(-4)
                         & 1.7(-3) & 3.2(-4)\\ 
16$\times$16$\times$4 & 32 & 2.3(-3) & 1.3(-3) & 2.2(-4) & 1.9(-3) & 3.0(-4)
                           & 7.8(-3) & 6.4(-4)\\ 
2$\times$2$\times$4 & 64 & 4.0(-4) & 4.1(-4) & 4.5(-4) & 6.1(-4) & 4.7(-4)
                         & 2.4(-4) & 2.0(-4)\\ 
4$\times$4$\times$4 & 64 & 4.2(-4) & 3.3(-4) & 2.6(-4) & 4.7(-4) & 2.8(-4)
                         & 2.7(-4) & 1.6(-4)\\ 
8$\times$8$\times$4 & 64 & 1.1(-3) & 6.6(-4) & 2.3(-4) & 9.6(-4) & 3.0(-4)
                         & 1.9(-4) & 3.2(-4)
\enddata
\end{deluxetable}

%-------------------------------------------------------------------------------
\clearpage

\begin{deluxetable}{rccccc}
\tablecolumns{6}
\tablewidth{0pc}
\rotate
\tablecaption{Gravitational Force Exerted by the Gas and Its Correlation Time\label{T:force}} 
\tablehead{
\colhead{Dimensions} & \colhead{Resolution}
& \colhead{$F_{x,\mathrm{rms}} / \xi m_p HP^{-2}$}
& \colhead{$F_{y,\mathrm{rms}} / \xi m_p HP^{-2}$}
& \multicolumn{2}{c}{$\tau_c / P$}\\
\colhead{[$H$]} & \colhead{[pt/$H$]} & \colhead{[10$^{-2}$]} & \colhead{[10$^{-3}$]}
& Exact (Eq.~\eqref{E:ct1}) & Estimated (Eq.~\eqref{E:ct2})}
\startdata
2$\times$2$\times$4 & 16 & 1.0$\pm$0.4 & \phantom{0}2.9$\pm$0.8
& 0.016 & 0.020\\ 
4$\times$4$\times$4 & 16 & 3.6$\pm$1.4 & \phantom{0}5.0$\pm$0.9
& 0.019 & 0.014\\ 
8$\times$8$\times$4 & 16 & 4.4$\pm$1.4 & \phantom{0}8.0$\pm$1.6
& 0.067 & 0.067\\ 
16$\times$16$\times$4 & 16 & 6.0$\pm$2.2 & 13.1$\pm$2.8
& 0.130 & 0.120\\ 
2$\times$2$\times$4 & 32 & 1.0$\pm$0.3 & \phantom{0}2.7$\pm$0.7
& 0.023 & 0.029\\ 
4$\times$4$\times$4 & 32 & 2.8$\pm$1.1 & \phantom{0}4.2$\pm$0.8
& 0.020 & 0.014\\
8$\times$8$\times$4 & 32 & 6.4$\pm$2.9 & \phantom{0}6.3$\pm$1.3
& 0.070 & 0.065\\
16$\times$16$\times$4 & 32 & 4.5$\pm$1.3 & 10.4$\pm$2.1
& 0.119 & 0.112\\
2$\times$2$\times$4 & 64 & 1.3$\pm$0.6 & \phantom{0}3.5$\pm$1.0
& 0.033 & 0.022\\ 
4$\times$4$\times$4 & 64 & 3.6$\pm$1.5 & \phantom{0}4.7$\pm$0.9
& 0.020 & 0.022\\ 
8$\times$8$\times$4 & 64 & 6.1$\pm$3.0 & \phantom{0}6.7$\pm$1.5
& 0.068 & 0.070
\enddata
\end{deluxetable}

%-------------------------------------------------------------------------------
\clearpage

%-------------------------------------------------------------------------------
\clearpage

%-------------------------------------------------------------------------------
\clearpage

%-------------------------------------------------------------------------------
\clearpage

%-------------------------------------------------------------------------------
\clearpage

%-------------------------------------------------------------------------------
\clearpage

%-------------------------------------------------------------------------------
\clearpage

%-------------------------------------------------------------------------------
\clearpage

%-------------------------------------------------------------------------------
\clearpage

%-------------------------------------------------------------------------------
\clearpage

%-------------------------------------------------------------------------------
\clearpage

%-------------------------------------------------------------------------------
\clearpage

%-------------------------------------------------------------------------------
\end{document}